# Avalanche-like plasticity in complex concentrated alloys: a review across scales

M. Knapek[1,2,*], T. Tayari[1], K. Máthis[1], and M. Janeček[1]

[1] *Department of Physics of Materials, Faculty of Mathematics and Physics, Charles University, Ke Karlovu 5, 121 16 Prague 2, Czech Republic*
[2] *Research Centre UNIZA, University of Žilina, Univerzitná 8215/1, 010 26 Žilina, Slovak Republic*

## ABSTRACT

Complex concentrated alloys (CCAs), including high- and medium-entropy alloys, deform in chemically heterogeneous energy landscapes where dislocation glide, solute aging, twinning, phase transformation and microstructural barriers may all contribute. This review discusses avalanche-like and serrated plasticity in CCAs across scales. The first part separates the relevant length scales and methods used to access them. At microscopic and mesoscopic scales, acoustic emission (AE) and microcompression studies reveal discrete dislocation avalanches and strain bursts that may be hidden in conventional macroscopic curves. At the specimen scale, local extensometry and digital image correlation (DIC) studies of the Portevin–Le Chatelier effect show how collective defect dynamics can organize into deformation bands and macroscopic stress serrations. Together, these approaches show that plastic flow may appear smooth only as an average response, while remaining intermittent at finer scales. The second part reviews direct CCA evidence, with emphasis on AE, stress-serration statistics, microplasticity, DIC, nanoindentation and small-scale deformation. A central conclusion is that serrated flow in CCAs should not be treated as a single phenomenon. Depending on chemistry, temperature, strain rate and microstructure, it may originate from dynamic strain aging, twinning, martensitic transformation, slip localization, or a combination of these mechanisms. Reported power-law-like distributions and exponents partly overlap with those known from simpler crystals and alloys, and the present evidence does not establish a distinct CCA-specific universality class. CCAs are therefore best viewed as tunable systems in which chemical disorder, short range order, phase stability and microstructure can modify the nucleation, arrest and synchronization of collective plastic events.

Address correspondence to E-mail: michal.knapek@matfyz.cuni.cz

## Introduction

In engineering practice, plastic deformation is usually characterized by macroscopic stress–strain curves, which often appear smooth. At the microscopic level, however, plastic flow proceeds through discrete defect-mediated processes. Mobile dislocations, twin boundaries or transformation interfaces may be temporarily impeded by obstacles such as solute atoms, precipitates, and other lattice defects, causing local stresses to build up. Once a barrier is overcome, the resulting motion redistributes stress and may activate neighboring defects. When such events become correlated, the resulting cascade of activity is referred to here as a plastic avalanche [1–3].

Avalanche-like plasticity is therefore not a separate microscopic deformation mechanism. Rather, it describes an intermittent and collective mode in which mechanisms such as dislocation slip, twinning or phase transformation may operate. Many small events occurring at different positions and times may overlap and produce an apparently smooth deformation curve at the macroscopic level. At finer scales, however, such activity can be detected as discrete AE bursts or local strain and displacement bursts, while sufficiently large or synchronized events may additionally become visible as deformation bands, nanoindentation pop-ins or macroscopic stress serrations [2, 4, 5]. These observables are not mutually equivalent because each represents a different spatially and temporally filtered projection of the underlying deformation activity.

This scale-dependent behavior is relevant to both fundamental and applied materials science. Measurements sensitive to intermittency can reveal irreversible microplastic activity before the conventional macroscopic yield point, probe the activation and evolution of weak sites within the initial microstructure, and show how initially distributed plasticity develops into strain localization [4, 6, 7]. Individual bursts become particularly important in micro- and nanoscale specimens, where a single event may account for a substantial fraction of the total accumulated strain [4, 8]. Large or synchronized events may also influence the reproducibility of yielding, surface quality and formability, particularly when deformation localizes into propagating or intermittently nucleating bands [5, 9]. Avalanche-resolved measurements therefore provide a bridge between defect-scale dynamics and specimen-scale mechanical response.

A practical starting point is provided by relatively simple metals. In highly anisotropic hexagonal close-packed (HCP) crystals such as ice, zinc and cadmium, AE measurements show separated dislocation avalanches with broad power-law-like energy distributions [3, 10, 11]. Face-centered cubic (FCC) metals such as copper show apparently similar behavior, but individual avalanches are less obvious in bulk tests because multiple slip systems and the development of dislocation substructure screen or interrupt long-range correlated motion more effectively [2, 3]. Small-scale testing makes the discrete response more visible: as the tested volume decreases, individual strain bursts contribute a larger fraction of the accumulated plastic strain, giving rise to the well-known "smaller is wilder" behavior [3, 4].

Conventional alloys add a second layer of complexity through time-dependent solute–dislocation interactions. In the dynamic strain aging (DSA) regime, mobile dislocations remain temporarily arrested at obstacles long enough to be additionally pinned by diffusing solutes. When the applied stress overcomes the strengthened obstacle, collective unpinning may produce abrupt strain bursts or stress drops. This is the physical basis of the Portevin–Le Chatelier (PLC) effect in many dilute alloys, which links solute kinetics, negative strain-rate sensitivity, localized deformation bands, AE activity and stress serrations [9, 12]. This classical example also illustrates an important distinction: DSA is a physical mechanism, whereas a stress serration is one possible macroscopic manifestation of the resulting instability under particular temperature, strain-rate and loading conditions [7, 13].

Complex concentrated alloys (CCAs)[1] represent a natural but more complicated extension of this problem. Instead of being based on one principal element with dilute additions, they were introduced through multicomponent and near-equiatomic alloy design concepts by Cantor et al. [14] and Yeh et al. [15] and were later consolidated into the broader CCA framework in comprehensive critical reviews [16, 17]. For the present topic, the important point is not configurational entropy by itself, but that defects may move through spatially heterogeneous energetic and kinetic landscapes [18, 19]. Local elastic response, generalized stacking fault energies, lattice friction, diffusion kinetics and chemical short range order (SRO) may vary with the local atomic configuration [20, 21]. Many CCAs additionally contain precipitates, dendritic segregation, phase boundaries or transformation-prone regions, introducing further barriers and internal length scales [9, 18]. These features can either interrupt and spatially distribute plasticity or synchronize defect activity by creating common conditions for arrest, stress accumulation and release.

[1] In this review, the term CCA is used in a broad sense for metallic alloys containing several principal elements at high concentrations, including medium-entropy alloys (MEAs), high-entropy alloys (HEAs) and related multi-principal component alloys (MPCAs). HEAs are commonly understood as alloys with five or more principal elements, often near-equiatomic or with each principal element typically in the range of about 5–35 at.%, whereas MEAs usually contain fewer principal elements or lower configurational entropy. MPCAs is a more composition-neutral term emphasizing alloy design based on several principal components rather than on a strict entropy threshold. These definitions are not fully rigid, but they are useful here because all these materials share a key feature: dislocation motion takes place in a chemically complex lattice rather than in a dilute-solute matrix.

Several CCA families show serrated flow resembling conventional PLC behavior, but similarity of the mechanical curves does not establish a common microscopic origin [9, 22]. Elevated-temperature serrations may be controlled by DSA, as supported in appropriate cases by their temperature–strain-rate window, negative strain-rate sensitivity or strain-rate jump response [7, 12, 13]. Intermittent deformation under other conditions may instead involve collective dislocation slip, twinning, martensitic transformation, deformation band propagation, local thermal effects or several concurrent processes [20, 23, 24]. This review therefore treats serrated flow as one possible macroscopic expression of intermittent plasticity, not as a mechanism by itself. The central question is how chemical complexity, crystal structure, phase stability, microstructure and testing conditions influence whether elementary deformation events remain distributed, become locally correlated or synchronize into specimen-scale instabilities [9, 25].

## Terminology

In this review, avalanche-like plasticity means intermittent collective defect activity detected through AE, strain bursts, displacement bursts, local strain-rate fluctuations or stress drops. The term describes the dynamical character of deformation and does not by itself identify the active microscopic carrier or mechanism [2, 3]. Serrated flow is used more narrowly when such intermittency becomes visible in the macroscopic mechanical curve, whereas microplasticity refers to irreversible defect activity before or near the conventional macroscopic yield point [4]. An alloy may consequently contain many small avalanches without visible serrations, while one macroscopic serration may be the synchronized outcome of numerous smaller events rather than one elementary avalanche [9, 25].

## Scope and literature-selection criteria

The need for this review is connected to a gap between two bodies of literature. Existing CCA serrated flow reviews summarize stress drop types, testing windows and microstructural variables [9], while broader reviews of intermittent plasticity discuss dislocation avalanches, microplasticity, AE and statistical-physics models [2–4]. These two discussions still meet only partially. A scale-bridging view is useful because CCA-specific features—chemical disorder, SRO, metastability and multiphase microstructures—can influence how elementary avalanches become arrested, redistributed or synchronized into mesoscopic localization and macroscopic serrations [26–29].

The literature selection used here is intentionally pragmatic rather than formally systematic. Papers were included when they report serrated flow, DSA/PLC, AE, nanoindentation pop-ins, strain bursts, digital image correlation (DIC) or local strain intermittency, martensitic or twinning-related bursts, or statistical analysis of deformation fluctuations in CCAs. Foundational work on simple crystals, conventional alloys and statistical models is used only where it helps to interpret CCA data. General CCA papers without a clear connection to intermittent plasticity, avalanche statistics, serrated flow, AE, DIC, twinning/transformation-induced plasticity (TWIP/TRIP) bursts or deformation localization are mostly not discussed in detail.

Broader CCA literature is used mainly as background to define the wider CCA field and terminology [16, 17] and to discuss concrete mechanisms or observations such as solid-solution strengthening models [30, 31], direct SRO and atomic-level complexity studies [26–28], lattice strain and deformation kinetics work [19, 32], early mechanical and serrated-flow observations in CoCrFeMnNi- and $Al_x$CoCrCuFeNi-type alloys [33, 34], and refractory-alloy design studies [35, 36]. Fatigue, high-temperature, high-rate and extreme-load-bearing studies provide useful boundary conditions rather than central evidence for intermittency [37–47].

The discussion below therefore remains focused on papers that elaborate directly on intermittent plasticity, serrated flow or related deformation instabilities.

An additional limitation is that the thermomechanical treatment and processing history of the reviewed CCAs are not treated here as independent organizing variables. This is a deliberate scope choice, because the available literature is highly heterogeneous and often does not permit a controlled separation of composition, microstructure, processing history and testing conditions. Nevertheless, this factor is not secondary: prior deformation, casting or solidification route, homogenization, annealing, recrystallization, precipitation or ordering, grain size, texture, chemical SRO, residual/internal stresses and defect density can all influence the active deformation mechanisms and the associated intermittent dynamics. Consequently, differences in serration behavior, AE response, event cutoffs or apparent power-law-like exponents should not be overinterpreted as purely compositional or CCA-specific effects.

## Outline

The review is organized as follows. Section "Concepts from simple metals and conventional alloys" establishes the conceptual and methodological framework, distinguishing experimental observables, the interpretation of power-law-like statistics, jamming and depinning, the nonequilibrium-thermodynamic perspective, mild-to-wild plasticity, DSA/PLC and microplasticity. Section "Why CCAs are a distinct test platform" explains why CCAs provide a particularly useful test platform for intermittent plasticity. Section "Intermittent plasticity in CCAs" reviews direct experimental and atomistic CCA evidence, including serration-based extreme event metrics, and synthesizes the active mechanisms from crystal structure and phase constitution perspectives. Section "Comparison of reported power-law exponents" compares reported power-law exponents across material classes and observables. Section "Synthesis: what CCA intermittency adds to avalanche plasticity" integrates the evidence across scales, highlights established trends and open problems, and discusses the transition from local intermittent dynamics to continuum response and the experimental requirements for future work.

## Concepts from simple metals and conventional alloys

### Experimental observables of intermittent plasticity

The experimental signatures used to investigate intermittent plasticity are related, but they are not interchangeable. Stress serrations, AE bursts, small-scale strain or displacement bursts, DIC strain localization and nanoindentation pop-ins represent different measured quantities, sampling volumes, temporal windows and boundary conditions [2, 4, 48]. The same deformation process may produce several of these signals, while similar-looking signals may originate from different microscopic processes. Their principal distinctions are summarized in Table 1.

These distinctions have several consequences. Temporal coincidence between two signals supports a physical connection but does not establish a one-to-one equivalence. One macroscopic stress drop may contain a cluster of many AE-detected events, whereas AE may reveal microplastic activity that remains invisible in the bulk stress–strain curve [2, 7]. DIC may resolve the movement of a deformation band while averaging over the rapid elementary events from which that movement is assembled [5, 48]. Likewise, a displacement burst in a micropillar or a pop-in during indentation is conditioned by local geometry and instrument response and cannot be equated directly with bulk stress drop magnitude [4, 21].

The event variable must therefore be defined separately for each method. AE event energy, stress drop magnitude, displacement increment and DIC-derived strain-rate excursion are different experimental observables and should not generally be treated as interchangeable [2, 55]. Nevertheless, different observables may reflect the same underlying deformation event and can correlate closely under suitable experimental conditions; for example, individual stress drops and AE events have been directly correlated in small-scale deformation experiments [50]. Their statistical distributions can therefore be compared meaningfully, but numerical exponents should be regarded as directly equivalent only when the relation between the observables, event definitions and relevant scaling laws is established [3, 56].

With these measurement-level distinctions established, avalanche-like plasticity can be examined statistically through distributions of event sizes, AE

**Table 1** Principal experimental observables used to investigate intermittent plasticity. The methods provide complementary but non-equivalent information

| Observable | Directly measured or reconstructed quantity | Principal information provided | Main limitations and experimental dependencies | Representative references |
|---|---|---|---|---|
| Macroscopic stress serration or load drop | Variation in force, load or stress recorded during a mechanical test | Reveals that plastic activity has become sufficiently large or synchronized to affect the coupled specimen-machine response; under PLC conditions, it may accompany deformation band nucleation or propagation | It is not an elementary defect event. Magnitude and morphology depend on loading mode, machine stiffness and feedback, specimen geometry, sampling frequency, filtering and detrending. Curve morphology alone does not identify the microscopic mechanism | [2, 7, 9] |
| AE burst | Transient electrical waveform recorded by an AE sensor in response to elastic waves emitted by a rapid, localized source event | Provides high temporal sensitivity to rapid irreversible processes, including microplastic events too small to affect the macroscopic curve; amplitude, duration, frequency content and signal energy can characterize changes in detected activity | AE does not directly measure plastic strain or the energy released at the source. The measured waveform is affected by propagation, attenuation, reflections, specimen geometry, source–sensor distance, sensor bandwidth, coupling, thresholding and event-identification procedures | [7, 23, 49, 50] |
| Strain or displacement burst in a small-scale mechanical test | Sudden displacement or strain increment under load control, or the corresponding load drop under displacement control | Measures the direct mechanical contribution of an intermittent event to deformation of a micro- or nanoscale specimen | Burst magnitude depends on specimen dimensions, free surfaces, machine compliance, loading mode, feedback response and acquisition rate. Focused-ion-beam preparation and finite-size effects may alter the observed event population | [4, 6, 50, 51] |
| DIC strain localization | Image sequence from which surface displacements are reconstructed; strain or strain rate is subsequently obtained through spatial differentiation | Reveals where deformation localizes and how heterogeneous strain fields or deformation bands nucleate, propagate, interact and arrest | DIC provides a spatially and temporally filtered surface projection, not the full three-dimensional internal field. Resolution depends on image scale, subset and step size, frame rate, correlation quality, smoothing and strain calculation procedure. DIC alone does not identify the active defect carrier | [5, 48, 52] |

**Table 1** continued

| Observable | Directly measured or reconstructed quantity | Principal information provided | Main limitations and experimental dependencies | Representative references |
|---|---|---|---|---|
| Nanoindentation pop-in | Sudden penetration-depth excursion under load control, or a corresponding load excursion under displacement control, in a load–depth record | Detects the onset or burst-like progression of highly localized irreversible deformation beneath the indenter | The subsurface stress field is heterogeneous and multiaxial. Pop-ins may reflect dislocation nucleation or multiplication, twinning, transformation, cracking or interfacial processes. Their magnitude also depends on indenter geometry, surface preparation, loading rate, drift and instrument compliance | [21, 53, 54] |

event energies, durations, displacement increments, stress drops or local strain-rate excursions [2, 3]. The physical meaning of the event variable remains method-specific: AE analysis may use measured event energy or squared event amplitude, whereas mechanical measurements may use stress drop magnitude, displacement increment or slip-step size [23, 55]. Detection thresholds, event-identification rules, machine stiffness, detrending procedures and spatial and temporal resolution determine which part of the underlying event hierarchy is observed [4, 56]. The interpretation of a power-law-like distribution and its exponent therefore requires the additional criteria outlined below.

## Interpreting power-law-like statistics

A broad, approximately linear region on logarithmic axes is often described as evidence of power-law behavior, $p(x) \propto x^{-\beta}$. Such a plot is useful for visualizing a heavy-tailed event population, but it may not be by itself sufficient to establish a power-law distribution. The event variable $x$, probability representation, lower cutoff $x_{min}$, fitted interval, number of events and any upper cutoff should be reported [57, 58]. Where the underlying data permit, exponent estimation should preferentially use unbinned maximum-likelihood methods, accompanied by goodness-of-fit evaluation and comparison with plausible alternatives such as an exponentially truncated power-law, lognormal, stretched-exponential or exponential distribution. Linear regression of a binned log–log histogram can provide a visual guide, but may bias both the fitted range and the exponent [57].

When statistically supported, a power-law-like distribution indicates that event magnitudes are broadly distributed and that no single characteristic value is evident within the fitted observational interval [3, 57]. This is physically informative because rare large events remain relevant, and controlled changes in the exponent, upper cutoff or event rate may reflect changes in the spatial extent, arrest or synchronization of plastic activity [55, 56]. A truncated power law is not simply a failed power law—its cutoff may contain information on specimen dimensions, machine stiffness, work hardening, microstructural obstacles or localization length scales [56]. Extreme event statistics may also have predictive value: the evolution of an upper cutoff or expected maximum event size has been used to estimate the largest serration reached before fracture in selected CCAs [59]. Such a quantity should, however, be interpreted as a system- and model-dependent extreme event metric rather than a universal failure criterion.

Several stronger conclusions do not follow from an event-size power-law alone. It does not establish self-organized criticality, identify a depinning or jamming mechanism, demonstrate a unique microscopic deformation carrier, or define a universality class [2, 3]. Similar exponents may arise from different mechanisms, whereas the same mechanism may yield different values when crystal orientation, loading rate, hardening, specimen size, event definition or detection threshold changes [55, 56]. Exponents obtained from AE event energy, stress drops, displacement bursts or DIC-derived variables are also not directly interchangeable because these observables involve different physical transfer functions and degrees of coarse graining. In CCAs, where dislocation slip, DSA, twinning, transformation and damage may coexist, statistical analysis should complement rather than replace mechanism identification [9, 23].

## Jamming, depinning and a nonequilibrium-thermodynamic perspective

Intermittent dislocation dynamics have been interpreted through several related but distinct statistical-mechanical pictures. In a jamming-dominated regime, resistance to plastic flow arises primarily from the evolving long-range interactions among dislocations themselves, which generate metastable configurations and internal stress fields without requiring a fixed external pinning landscape. Discrete dislocation dynamics (DDD) simulations of nominally pure systems have consequently shown broad avalanche statistics and an extended critical-like response over a range of applied stresses rather than only at a single tuned critical stress [60, 61]. In a depinning-like regime, by contrast, dislocation motion is controlled more strongly by a spatially heterogeneous population of relatively persistent obstacles. Collective escape from this obstacle landscape may then become concentrated near a critical driving stress. Increasing the strength or density of quenched obstacles can therefore shift a system from dislocation–dislocation interaction-dominated jamming toward obstacle-dominated pinning/depinning behavior [62, 63]. Jamming and depinning should consequently not be used as synonyms, and an event-size exponent alone cannot establish which regime applies.

A recent complementary description has been proposed in the nonequilibrium-thermodynamic framework of McDowell and Liu and its extension to avalanche plasticity [64, 65]. McDowell and Liu formulate thermally activated dislocation plasticity as evolution through constrained local equilibrium states of interacting subsystems, such as grains or phases. Local defect configurations and internal stresses enter as internal state variables, while thermal and configurational entropy changes accompany irreversible transitions between states. Reaction probabilities are distinguished from thermodynamic driving forces, and intrinsic entropy production provides a measure of irreversible ensemble evolution, with the framework intended ultimately to support reduced-order internal state variable descriptions of crystal plasticity [64].

McDowell extends this construction specifically to avalanche plasticity by introducing the degree of correlation of precursor dislocation-barrier reactions as a descriptor of their collective organization [65]. Within this formulation, relatively decorrelated processes such as dislocation multiplication, cross-slip and reactions with weak barriers dominate at earlier stages, whereas the development of internal stress fields during jamming increases the degree of correlation as the system evolves toward a depinning regime. The power-law-like regime is associated with a plateau in this degree of correlation and configuration-dependent scaling is attributed to the jamming transition, whereas universal scale-free scaling up to a finite maximum avalanche size is associated specifically with the depinning transition. Within this description, the evolution of the internal defect state, reaction correlations and associated entropy production provides a thermodynamic measure of the irreversible organization of the deforming system [64, 65].

This thermodynamic construction provides a useful and physically motivated bridge between defect reactions, statistical intermittency and continuum internal state descriptions. As a relatively recent theoretical synthesis, however, its broader experimental validation and general applicability to avalanche plasticity are still developing. Earlier DDD studies show that scale-free or extended critical-like behavior can also arise in jamming-dominated dislocation systems without quenched pinning, and that the statistical response evolves as external disorder becomes more important [60–63]. The degree-of-correlation framework and the distinction between jamming and depinning therefore offer valuable additional tools for interpreting avalanche behavior, while their identification should be based on converging mechanistic evidence rather than on a power-law-like event distribution alone.

## Mild-to-wild plasticity and scale dependence

A complementary description of intermittent plasticity is provided by the "mild-to-wild" framework, which compares materials according to how plastic strain is distributed among deformation events [66]. In mild plasticity, deformation proceeds through many small, overlapping events yielding a nearly smooth macroscopic response where no single burst dominates the accumulated strain or released AE energy. In wild plasticity, a substantial fraction of strain or AE energy is released through intermittent bursts with heavy-tailed statistics, so rare large events remain important. HCP ice, zinc and cadmium are paradigm cases because strong plastic anisotropy restricts active slip, limits dislocation network formation and leaves long-range elastic interactions only weakly screened [3, 10, 11, 66]. FCC and body-centered cubic (BCC) metals usually exhibit milder plasticity since dislocation multiplication, forest hardening and cell formation reduce the mean free path of correlated motion, so fluctuations become more bounded and macroscopic flow appears smoother [3]. The transition is not controlled by crystal structure alone. It also depends on the ratio between external sample size and an internal length related to the dislocation structure. When the sample size approaches this internal length, as in micropillars, individual bursts are no longer averaged out and plasticity becomes visibly intermittent [51, 67–72]. This mild-to-wild transition is illustrated in Fig. 1 for compressed Mo micropillars. As the pillar diameter decreases, plastic flow changes from a milder, more distributed response to scale-free-like burst statistics and finally to supercritical, often system-spanning events. This simple reference case is useful before turning to CCAs, where chemical disorder, SRO and phase heterogeneity add further internal length scales and obstacles.

Alloying may either suppress or amplify intermittency, depending on how it modifies the defect landscape. Quenched disorder, solutes, precipitates, grain boundaries and phase interfaces can interrupt long slip events, shorten the mean free path of mobile dislocations and reduce the largest avalanches. In this sense, obstacles may "tame" wild plasticity by fragmenting collective slip into smaller events. Zhang et al. demonstrated this idea directly in Al and Al-alloy micropillars, where stronger solute-cluster or precipitate pinning shifted the wild-to-mild transition to smaller pillar diameters and led to the useful principle that "dirtier is milder" [73].

## Dynamic strain aging and macroscopic serrated flow

A different route to amplified intermittency arises from spatially heterogeneous or time-dependent pinning. Such pinning can store elastic energy and release it abruptly once local barriers are overcome. DSA is the classical example: when the waiting time of temporarily arrested dislocations becomes comparable to the diffusion time of solute atoms, additional solute pinning develops during deformation. This produces a negative strain-rate sensitivity window, promotes strain localization, and can lead to deformation bands and PLC serrations [74, 75].

The classical DSA/PLC framework is summarized schematically in Fig. 2. Negative strain-rate sensitivity arises when solute pinning competes with dislocation motion, producing a regime in which homogeneous flow becomes unstable. Depending on strain rate and temperature, this instability appears macroscopically as type-A, type-B or type-C serrations, which reflect different modes of deformation band propagation.

The conventional PLC effect also shows why the morphology of the mechanical curve should not be treated as a mechanism by itself. Type-A, type-B and type-C serrations are commonly linked to different strain-localization modes: continuously propagating deformation bands, hopping or relay-like bands, and more spatially uncorrelated band nucleation, respectively [74–77]. This classification is useful because it connects the macroscopic stress record with the spatiotemporal character of band motion. However, it remains phenomenological: the apparent serration type can also depend on strain rate, temperature, work-hardening rate, machine stiffness, sample geometry and measurement resolution. Moreover, a stress drop recorded by the testing machine is already a collective, coarse event. It may reflect band nucleation or propagation, but the band itself can involve many smaller dislocation avalanches. Therefore, macroscopic stress drop statistics do not necessarily match the statistics of elementary dislocation events detected by AE, nor do they directly identify the microscopic origin of the instability [78].

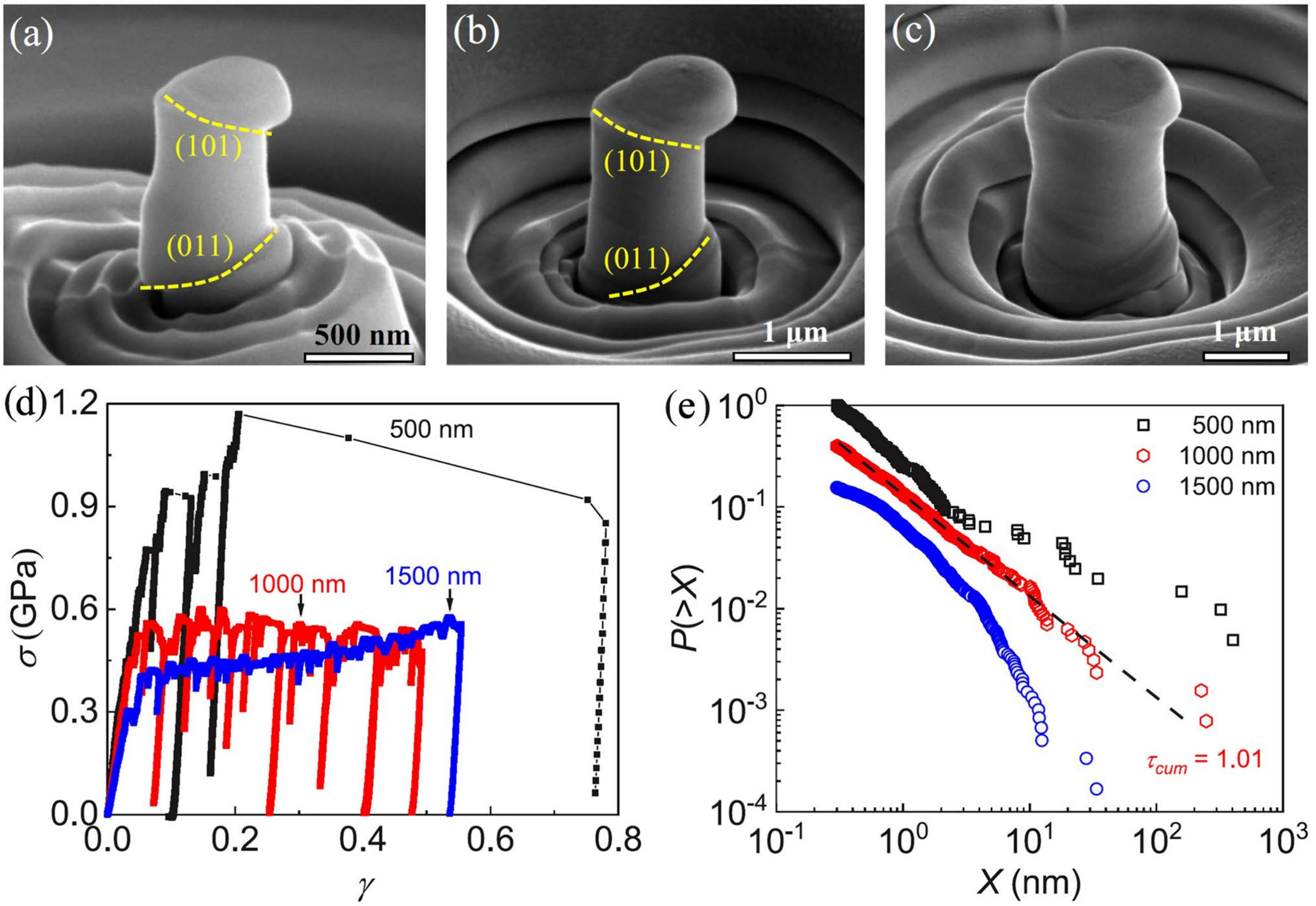

**Figure 1** Size-dependent intermittent plasticity and mild-to-wild transition in small-scale Mo micropillars. **a–c** SEM images of [112]-oriented Mo micropillars with diameters of **a** 500 nm, **b** 1000 nm and **c** 1500 nm after compression. The marked slip traces in **a,b** illustrate the local single-slip character of plastic flow in the 500 nm and 1000 nm pillars. **d** Representative intermittent shear stress–strain curves for the three pillar sizes. **e** Cumulative distributions of plastic displacement bursts, *P*(>*X*), detected during loading. The dashed line is a power-law guide fitted to the 1000 nm pillar data. The 1000 nm pillar shows scale-free-like burst statistics, whereas the larger 1500 nm pillar exhibits a milder, more distributed response and the smaller 500 nm pillar exhibits supercritical, often system-spanning displacement bursts. (Adapted from [71] with permission).

Several practical consequences follow from this discussion before turning to CCAs. Macroscopic stress serrations are coarse signals: each stress drop may contain many elementary events, and its magnitude also depends on machine stiffness, specimen geometry and strain localization [2, 9]. Irreversible microplastic events may appear well before the conventional yield point, implying that the apparently elastic part of a stress-strain curve can still contain defect activity [4, 7]. Finally, as discussed above, similar heavy-tailed statistics may arise from different physical mechanisms, particularly when the measured signal is an indirect or coarse-grained projection of the underlying deformation [3, 23].

The wider literature on crackling noise is useful here mainly as a conceptual reference frame for intermittent response in slowly driven systems [79–81]. In plasticity, the jamming and depinning pictures are related but distinct, as discussed in Section "Jamming, depinning and a nonequilibrium-thermodynamic perspective": jamming emphasizes self-generated arrest within an interacting dislocation network, whereas pinning/depinning emphasizes motion controlled by a persistent heterogeneous obstacle landscape [60, 62, 63]. For alloys, and especially CCAs, this statistical-mechanical language must be combined with the more specific frameworks of DSA, PLC instability and deformation band propagation [74–77]. These concepts are therefore used as reference frames, not as evidence that every CCA stress drop has the same microscopic origin.

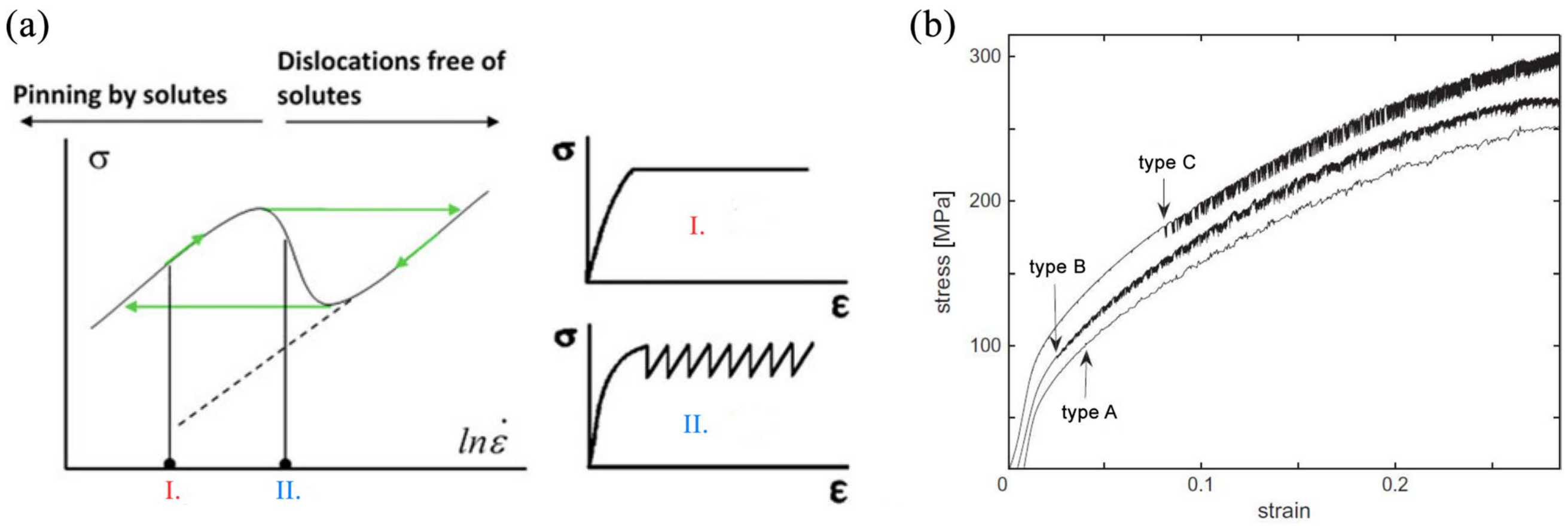

**Figure 2** Classical dynamic strain aging and Portevin–Le Chatelier instability. **a** Schematic transformation of a monotonic stress–strain-rate relation into an N-shaped curve when solute pinning produces negative strain-rate sensitivity. The sketches on the right illustrate smooth flow outside this regime and serrated flow within it. **b** Representative tensile curves of an Al–Mg alloy at room temperature showing type-A, type-B and type-C PLC serrations at decreasing imposed strain rate. Arrows indicate the critical strain for the onset of instability. (Adapted from [74]).

## Microplasticity and complementary experimental evidence

Intermittent plasticity is not restricted to the macroscopically plastic regime. Microplasticity refers to the first irreversible defect activity occurring below, or close to, the conventional macroscopic yield point and therefore provides an especially sensitive probe of weak sites in the initial microstructure [4]. Maaß and Derlet emphasized that, although this regime may appear homogeneous when averaged over a bulk specimen, the earliest irreversible events are inherently local and originate from the most easily activated sources or obstacles [4].

Classical descriptions emphasized internal friction, anelastic recovery, or dislocation segments breaking away from weak obstacles; modern small-scale testing reframes the same regime statistically as the earliest localized sampling of microstructural weak spots. In a CCA, these sites may be dislocation sources, grain-boundary stress concentrations, local chemical motifs, dendritic fluctuations, precipitate/matrix interfaces or regions of locally low fault energy. Recent microplasticity work also demonstrates why similar-looking power laws should not be regarded as universal fingerprints. Sparks and Maaß found that scaling exponents in nominally simple single crystals vary with structure, orientation and driving conditions [55]. Rizzardi et al. showed that alloy microstructures can shift gradually from scale-free to scale-dependent microplasticity as obstacles change from shearable to incoherent [56]. These results are a useful warning for CCAs: composition alone is not enough to interpret an exponent.

Loading–unloading experiments can serve as a useful diagnostic tool. The Kaiser effect denotes a memory-like AE response in which the material remains largely acoustically silent during reloading until the previously reached maximum stress, load or strain is exceeded. AE activity therefore reappears mainly when new irreversible defect activity is triggered beyond the prior maximum, rather than by simply repeating the same elastic loading path [82]. In plasticity studies this behavior is not merely a technical detail; it is consistent with partial exhaustion or rearrangement of the most easily activated sources during the previous loading step.

A final lesson is that spatial and microstructural information matter. AE has excellent temporal resolution and can detect very small, rapid events, but identifying the source mechanisms accurately might be difficult. Slip, twinning, phase transformation or cracking may all generate burst-like signals which can be concurrently active, thus causing complex AE response. DIC has coarser temporal resolution, but it shows where strain localizes and how deformation bands nucleate, propagate or arrest. Diffraction and electron microscopy add the missing structural information by identifying whether the active carrier is dislocation slip, twinning, martensitic transformation or another defect process. A convincing interpretation of intermittency in CCAs should therefore combine statistics with spatial localization and microstructural evidence whenever possible, rather than relying on the stress curve alone [4, 9, 11, 74, 75].

## Why CCAs are a distinct test platform

For avalanche plasticity, the key feature of CCAs is not configurational entropy alone, but the spatially heterogeneous energy landscape through which defects move. Local lattice distortion, elastic misfit and chemical fluctuations can alter dislocation line energies, local Peierls barriers and depinning stresses. Local chemistry and SRO can also modify generalized stacking fault energies and thereby shift the balance between full dislocation slip, partial slip, twinning and deformation-induced transformations [26–28, 30, 31]. Multiphase or dendritic microstructures introduce additional obstacles and internal stress fields. This distinction matters because many descriptors used for CCAs are mechanistically too broad for avalanche analysis. Terms such as chemical disorder, lattice distortion or sluggish diffusion become useful only when linked to specific deformation variables, such as obstacle strength, mobile dislocation dynamics, strain-rate sensitivity, localization length scales and the statistical distribution of deformation events. Configurational entropy helps define the alloy concept, but intermittency is more directly affected by local lattice resistance, fault energy fluctuations, solute atmospheres, phase boundaries, precipitates, twin boundaries and the density of mobile dislocation sources. Some of these features act as quenched obstacles, whereas others evolve during deformation or thermal exposure.

The same CCA landscape can both randomize and synchronize plasticity. Randomization occurs when local variations in chemistry or microstructure interrupt slip propagation and distribute plastic activity over many weak sites [26–28, 30, 31, 73]. Synchronization occurs when many mobile dislocations are arrested by similar barriers or solute atmospheres and subsequently released collectively, producing AE bursts or macroscopic serrations [74–78, 83, 84]. This dual role is the main reason why CCAs should not be regarded simply as more disordered dilute alloys [9, 29]. At present, the available evidence does not justify the strong claim that CCAs form a new universality class of avalanche plasticity [55, 56, 60, 85–88]. A safer statement is that CCAs broaden the range of tunable control parameters—chemical disorder, SRO, local fault energy, metastability and phase constitution—and therefore provide a useful platform for testing how avalanche statistics evolve with microstructural complexity.

Several types of evidence support this cautious view. DSA-related strengthening and negative strain-rate sensitivity have been reported in Al-bearing and chemically modified CCAs [89–91]. The lattice-strain framework and related deformation studies show that dislocation kinetics in CCAs can deviate strongly from those of conventional FCC alloys even when macroscopic flow is not obviously serrated [19, 32]. At the same time, recent SRO studies suggest that local order can modify defect pathways and incipient plasticity, but they do not yet provide a simple one-to-one link between SRO and avalanche exponents [26–28, 92].

The useful question is therefore not whether CCAs are fundamentally different in all respects; they are not. Rather, it is which CCA-specific variables change the probability that elementary events remain distributed, become locally correlated, or synchronize into a macroscopic instability. This question guides the discussion of the direct evidence below.

## Intermittent plasticity in CCAs

### Experimental evidence across scales

The equiatomic FeAlCrMo alloy provides one direct CCA example in which macroscopic stress serrations, negative strain-rate sensitivity and AE activity were examined within the same DSA/PLC framework. Knapek et al. reported serrated flow during compression of equiatomic FeAlCrMo at elevated temperatures [13]. The alloy featured a BCC solid solution, while the as-cast microstructure retained dendritic chemical heterogeneity, including Mo-rich regions. This combination is useful: the alloy is structurally simple enough to discuss as a model BCC CCA, yet it is not chemically homogeneous on the mesoscale. The serration morphology in FeAlCrMo changes systematically with deformation conditions. At 400 °C, decreasing the strain rate from $10^{-2}$ to $10^{-4}$ $s^{-1}$ changed the response from type-B-like serrations to larger C-type drops, and then to an atypical bimodal pattern with large drops separated by smaller serrations. Strain-rate-jump testing confirmed negative strain-rate sensitivity. These observations support a DSA/PLC interpretation over the description based solely on curve appearance.

Tayari et al. extended the same alloy from macroscopic serrations to AE-resolved microplasticity [7]. During compression at 300 and 400 °C, stress drops correlated with AE count rate, event amplitude and median-frequency changes. At 300 °C the response resembled B-type PLC behavior, whereas at 400 °C large C-type drops of about 20 MPa were accompanied by smaller intermediate serrations of roughly 2–3 MPa. This already suggests that more than one event scale is active during plastic flow. The most important observation for the present review is that AE activity was detected already in the macroscopically elastic or quasi-elastic region. Cyclic loading at 400 °C showed Kaiser-effect-like behavior: AE reappeared only when the previous maximum stress was reached. This behavior is consistent with the activation, exhaustion or rearrangement of mobile defect sources before conventional macroscopic yield. FeAlCrMo therefore connects classical DSA/PLC with microplastic avalanche activity in one material.

The same work reported power-law-like distributions of AE event energies (approximated as squared amplitudes $A^2$) with the $\beta$ exponents of about 1.4 for plastic events at 300 °C, 1.6 for plastic events at 400 °C and 1.7 for microplastic events at both temperatures [7]. These values are close to those reported for several simpler crystalline systems [3, 11], but the physical context is richer because DSA synchronization and chemical complexity are present at the same time.

The FeAlCrMo case is summarized in Fig. 3. It combines the original observation of unusual strain-rate-dependent serrated flow and stress drop/AE correlations with later AE evidence for microplastic activity and power-law-like event statistics. The example is useful because it connects DSA/PLC-type macroscopic instability with AE-detected intermittent plasticity in both the nominally elastic and plastic regimes of the same CCA system.

Al-containing CCAs provide the broadest serrated-flow dataset. The review by Brechtl et al. shows how composition, microstructure, loading mode, temperature, strain rate and serration type all influence the instability [9]. In $Al_{0.5}$CoCrCuFeNi, Brechtl et al. combined compression testing with statistical, dynamical and multifractal analyses [22]. Type-C serrations were associated with lower sample entropy, wider multifractal spectra and a higher probability of large stress drops, whereas Type-A and Type-B serrations showed higher complexity and a lower probability of large drops. This suggests that serration type can be a dynamical fingerprint of the evolving microstructure, not just a curve label.

The broader CCA literature summarized by Brechtl et al. also shows that serrated flow should not be reduced to a single Al-solute mechanism [9]. Al-containing CoCrFeNi-based alloys are prominent because Al additions can promote DSA-active behavior and may encourage ordering or precipitation [93–95]. A particularly direct illustration is provided by the comparison between CoCrFeNi and $Al_{0.5}$CoCrFeNi, where the Al-containing alloy exhibits much more pronounced serrations than the Al-free alloy [95]. A simple composition-level comparison is shown in Fig. 4. Under identical tensile conditions, CoCrFeNi deforms almost smoothly, whereas $Al_{0.5}$CoCrFeNi exhibits clear stress serrations. This supports the interpretation that Al can promote DSA/PLC-type instability in this alloy family, although the effect should be understood together with the accompanying changes in phase constitution and dendritic/interdendritic chemical partitioning. However, C additions can also produce room-temperature serrations in Cantor-type alloys [96], while other reports involve dendrite boundaries, sigma-phase particles, $L1_2$ or B2 ordering, mechanical twinning and cryogenic transformations [9], as will be particularized later.

At this point, the key question is whether macroscopically smooth flow is genuinely homogeneous or only appears homogeneous after spatial and temporal averaging. Lebyodkin et al. addressed this issue in FeCoNiTiAl over the range 200–700 °C by analyzing fine stress fluctuations using Fourier spectra, refined composite multiscale entropy and multifractal methods [25]. As summarized in Fig. 5, the alloy showed apparently smooth flow at 200, 600 and 700 °C, type-A serrations at 300 and 400 °C, and type-C serrations at 500 °C. After detrending, even the macroscopically smooth or pre-serration regimes contained non-random, temperature-dependent fluctuations, although their amplitudes were about two orders of magnitude smaller than visible PLC serrations [25]. This observation decouples collective dislocation dynamics from the mere presence of macroscopic stress drops: a smooth stress–strain curve may still contain signatures of self-organized plastic activity if recorded with sufficient resolution and analyzed appropriately. FeCoNiTiAl therefore provides a useful bridge between classical PLC studies and the broader problem of hidden intermittency in apparently stable plastic flow.

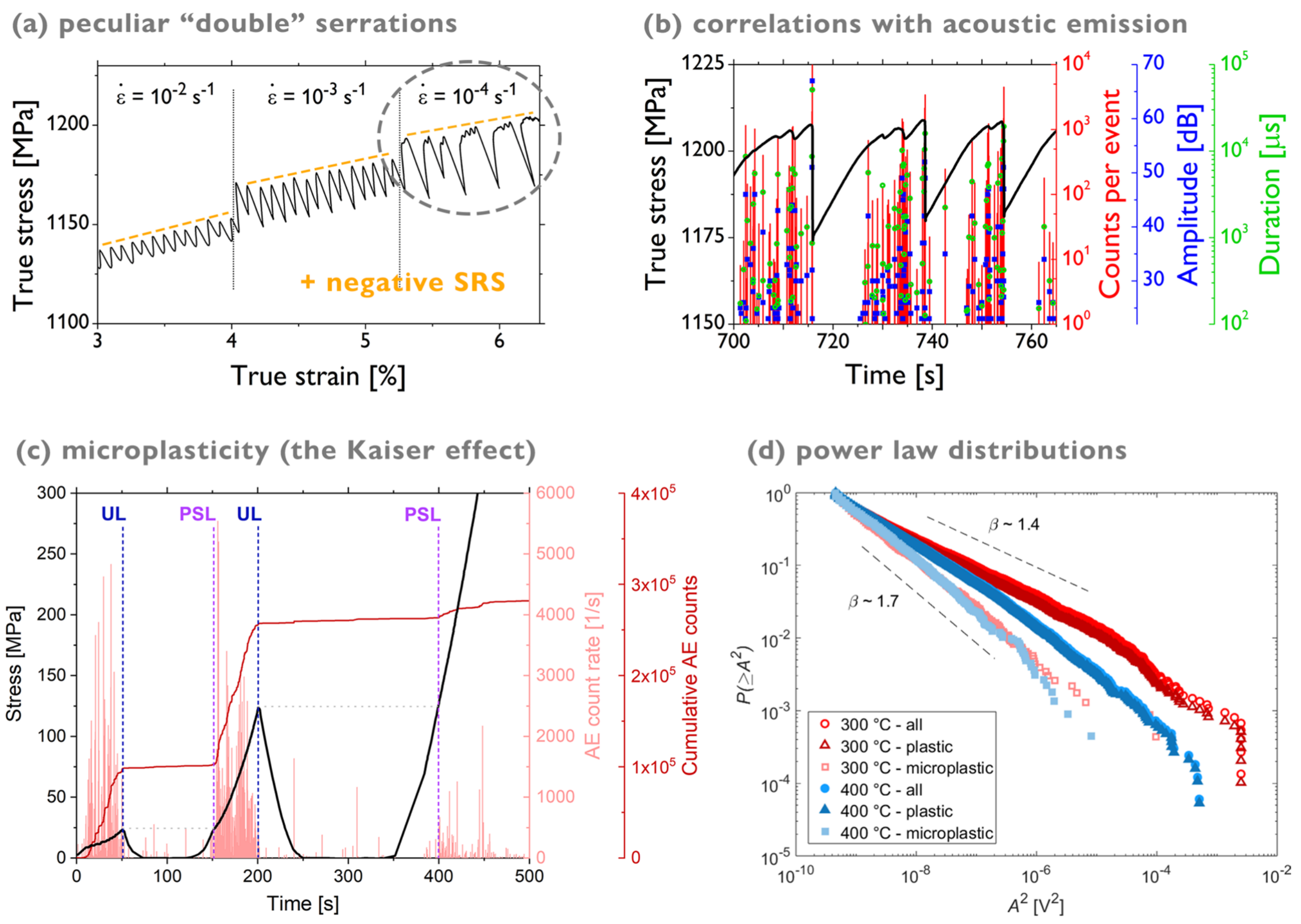

**Figure 3** Serrated flow, AE and microplastic intermittency in as-cast FeAlCrMo CCA. **a** Strain-rate jump test during compression at 400 °C showing strain-rate dependence of serrated flow, negative strain-rate sensitivity (SRS), and the development of peculiar double serrations at $10^{-4}$ $s^{-1}$. **b** Correlation between stress drops and AE activity during compression at 400 °C and $10^{-4}$ $s^{-1}$, shown through AE counts per event, event amplitude and duration. **c** Cyclic loading at 400 °C showing AE activity during early loading and its reappearance mainly after the previously reached stress level (PSL) is exceeded, consistent with Kaiser-effect-type microplasticity; UL denotes unloading. **d** CCDFs of squared AE event amplitudes, $A^2$, for microplastic and plastic regimes at 300 and 400 °C, showing power-law-like behavior over several orders of magnitude. (Panels (a,b) are adapted from [13] and panels (c,d) from [7].

A similar conclusion was drawn for the $Al_{0.3}$CoCrFeNi alloy, which was studied using DIC, AE and nonlinear analyses [5]. Although the alloy deformed macroscopically smoothly at room temperature, DIC maps and stress/AE analyses revealed transitions between distinct dynamical regimes, including blue-noise-like and reddened behavior. As illustrated in Fig. 6, the global stress–time curve can remain smooth while local strain-rate activity evolves heterogeneously along the gauge length. The purpose of this work was therefore not to identify another serrating CCA, but to expose the mesoscopic scale at which dislocation self-organization becomes visible even when the global stress–strain curve remains uniform. The study is particularly useful because it combines temporal and spatial information: DIC reveals mesoscale strain localization, whereas AE records faster burst-like activity. These observables do not measure the same events, but together they show how local plastic activity can be filtered into mesoscale strain patterns and, only under some conditions, into macroscopic serrations.

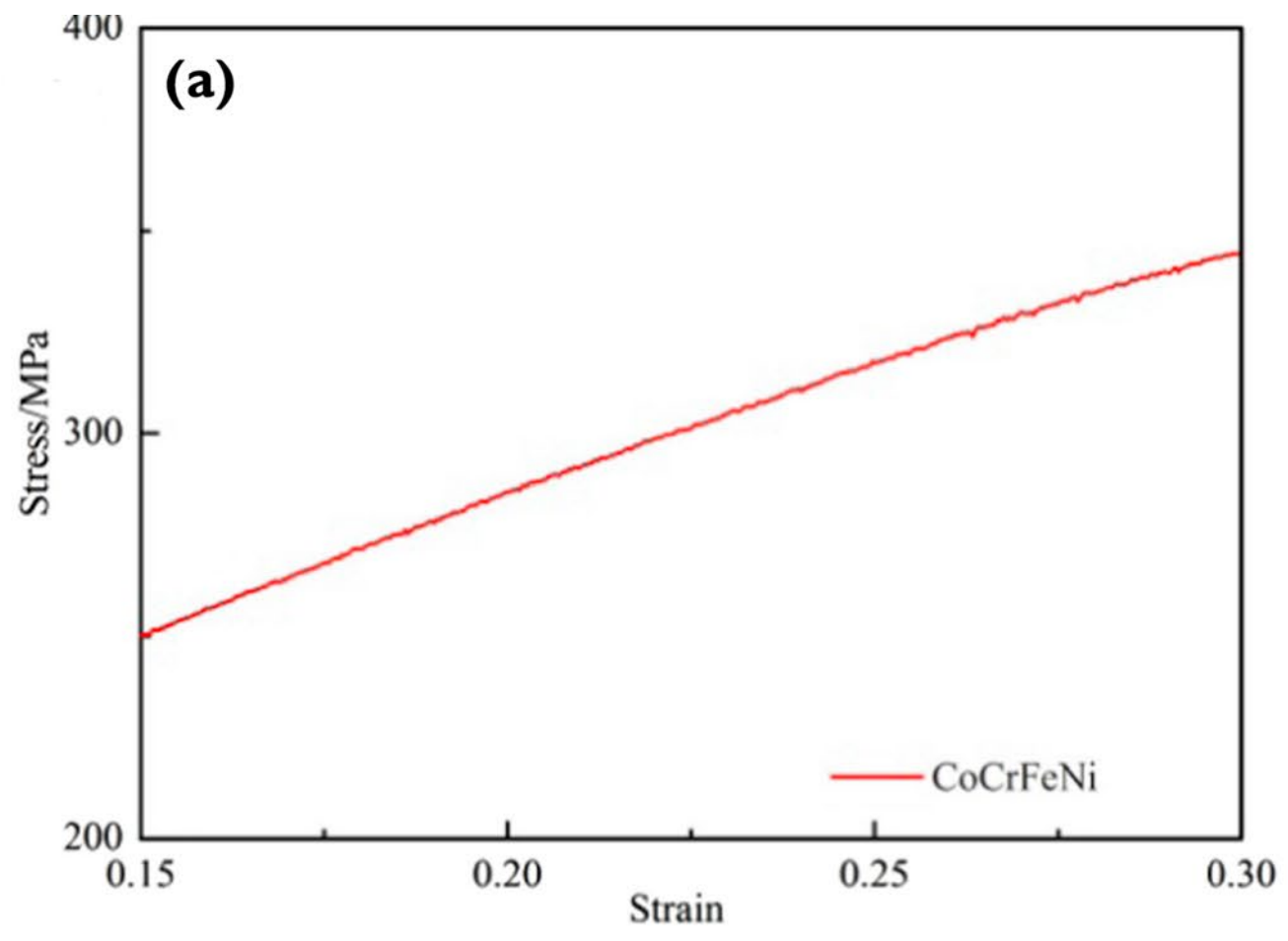

**Figure 4** Effect of Al addition on serrated flow in CoCrFeNi-based alloys. Engineering stress–strain curves at 400 °C and strain rate of $10^{-3}$ $s^{-1}$ for **a** CoCrFeNi and **b** $Al_{0.5}CoCrFeNi$. The Al-free alloy deforms almost smoothly, whereas the Al-containing alloy exhibits pronounced serrations, illustrating how alloying can shift a CCA into a DSA/PLC-susceptible regime. (Adapted from [95] with permission).

Small-scale CCA experiments provide a complementary view of elementary slip avalanches. Hu et al. correlated slowly compressed $Al_{0.1}CoCrFeNi$ nanopillars with in situ TEM observations and showed that large displacement bursts are linked to dislocation-band propagation, pile-up release and avalanche-like slip events [97]. Rizzardi et al. monitored slip-line formation associated with individual avalanches in a CCA, distinguishing events that reactivate existing slip traces from those that create new ones [6]. Nanoindentation studies on Al-containing CCAs and statistical models of incipient CCA plasticity further show that pop-in behavior depends on composition, loading rate and local chemical inhomogeneity [53, 54]. These studies also clarify why CCAs are attractive for microplasticity research. In small volumes, the number of active dislocation sources is limited and individual events become resolvable. Chemical disorder can increase the activation stress, but it may also fragment long slip paths and suppress system-spanning bursts. Heat treatment or composition series in micropillars could therefore provide a useful bridge between controlled small-scale intermittency and the more chemically complex disorder present in CCAs.

Cryogenic serrations require separate treatment. Classical DSA requires solute diffusion on the time-scale of dislocation waiting, which is generally incompatible with very low temperatures. Serrated flow at 4.2 K or in the liquid-nitrogen range in CrMnFeCoNi, CoCrFeNi and $FeCoCrNiMo_{0.2}$-type alloys has therefore been attributed to mechanisms such as sudden dislocation depinning, twinning, transformation, thermal softening or adiabatic heating rather than ordinary PLC aging [98–101]. These phenomena are relevant to avalanche plasticity, but their rate-controlling processes differ from elevated-temperature PLC behavior.

Recent papers broaden the discussion even further. AE analysis of $Fe_{40}Mn_{40}Co_{10}Cr_{10}$ reveals two avalanche branches associated with dislocation motion and twinning/detwinning, rather than a single event population [23]. As illustrated in Fig. 7, these two populations separate in the energy–amplitude relation and also show different maximum-likelihood energy exponent trends, demonstrating that a single AE dataset can contain superimposed deformation mechanisms. This is important for CCA studies because an apparent power-law distribution may represent a mixture of physically distinct event classes rather than one universal avalanche process.

A high-entropy shape-memory single-crystalline $Cr_{20}Mn_{20}Fe_{20}Co_{34.4}Ni_{5.6}$ shows intermittent AE during Lüders band-type martensitic transformation and thermal recovery [24]. In CrCoNi-based concentrated

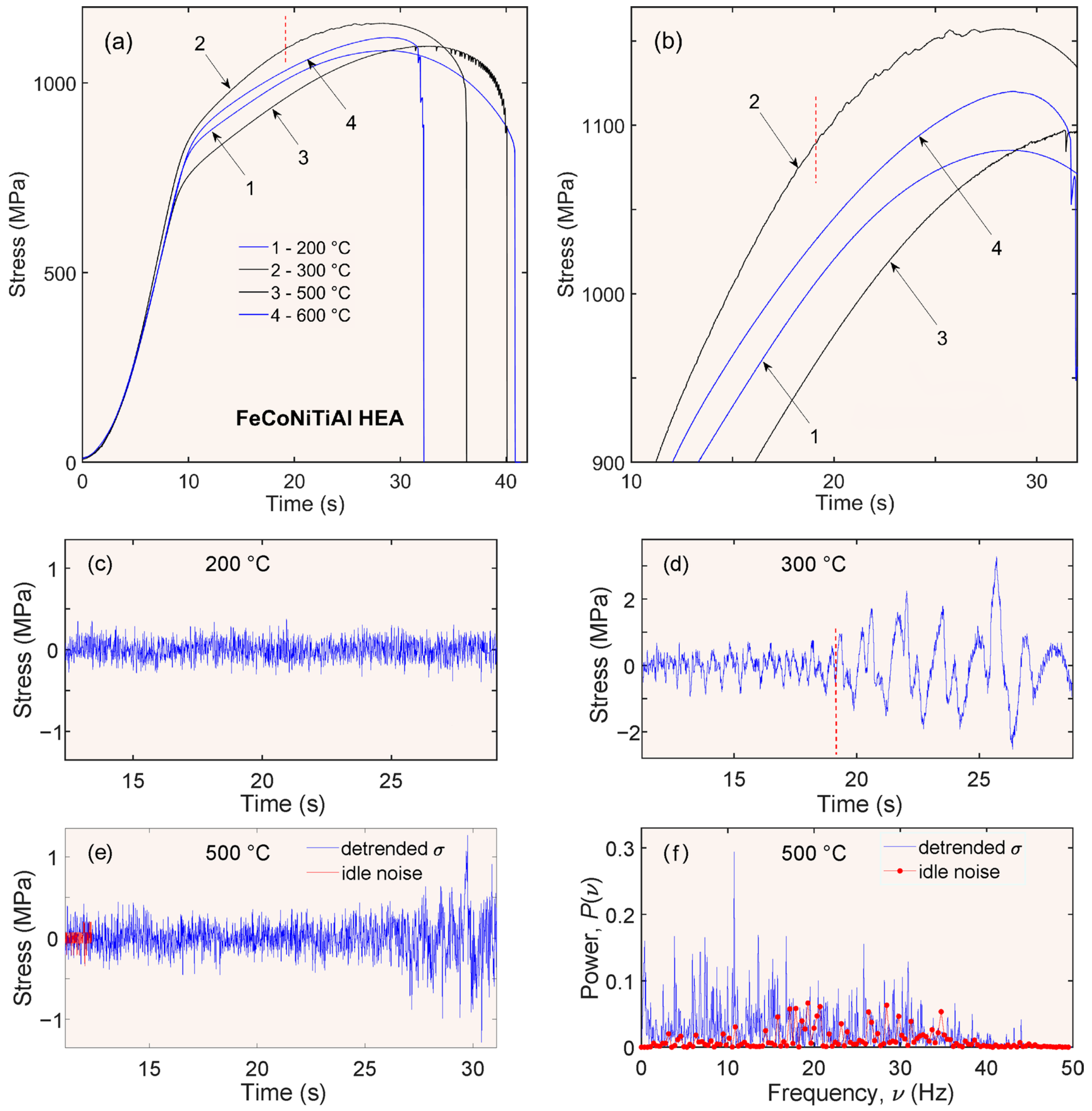

**Figure 5** Smooth and jerky flow in the FeCoNiTiAl alloy. **a** Representative stress–time curves recorded at 200, 300, 500 and 600 °C, showing the transition from macroscopically smooth flow to PLC-type serrated flow. **b** Magnified view of the deformation curves, highlighting type-A serrations at 300 °C and the onset of type-C serrations at 500 °C; the dashed vertical line marks the critical strain for the onset of visible instability. **c–e** Representative detrended stress signals: small fluctuations around apparently smooth flow at 200 °C, type-A serrations at 300 °C and stress fluctuations before the first type-C serration at 500 °C. The red trace in **e** shows the idle noise level for comparison. **f** Power spectral density of the detrended 500 °C signal, demonstrating that the fluctuations exceed the idle noise and contain structured temporal information. (Adapted from [25]).

alloys, multiscale observations distinguish confined and extended slip bands and relate their evolution to source operation, deactivation and reactivation [102]. Atomistic simulations provide complementary evidence for dislocation-avalanche serrations outside the DSA framework [103], as discussed in Section

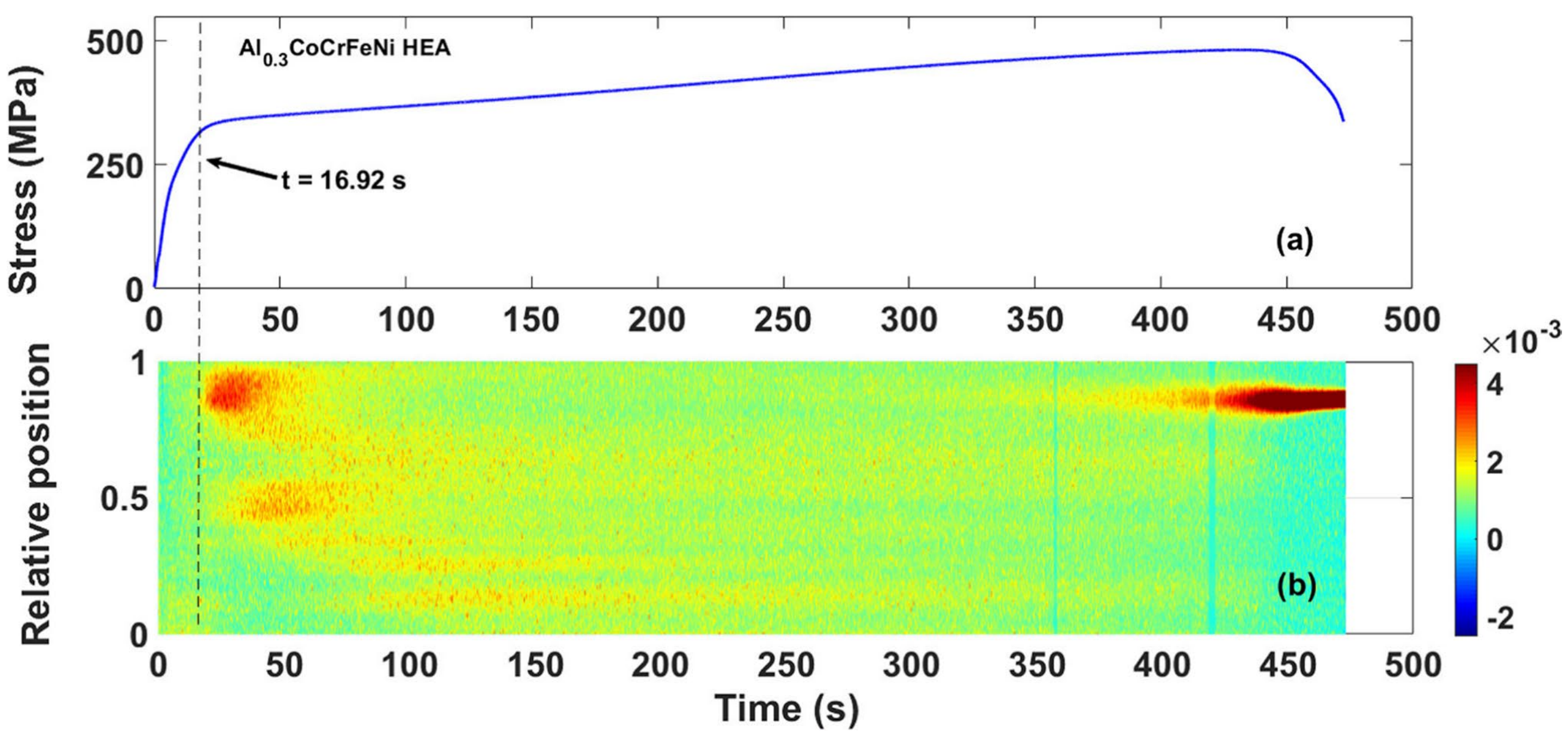

**Figure 6** Mesoscopic strain heterogeneity during macroscopically smooth plastic flow in $Al_{0.3}$CoCrFeNi CCA. **a** Stress–time curve recorded during room-temperature tensile deformation at $10^{-3}$ $s^{-1}$. **b** Corresponding DIC spatiotemporal map of local strain-rate activity along the tensile axis, with the vertical coordinate given as relative position along the gauge length. The dashed line at $t$=16.92 s marks the transition from a relatively uniform early plastic flow regime to macroscopically developed elasto-plastic deformation. Although the macroscopic curve remains smooth, the DIC map reveals localized and evolving mesoscale strain heterogeneity. (Adapted from [5] with permission).

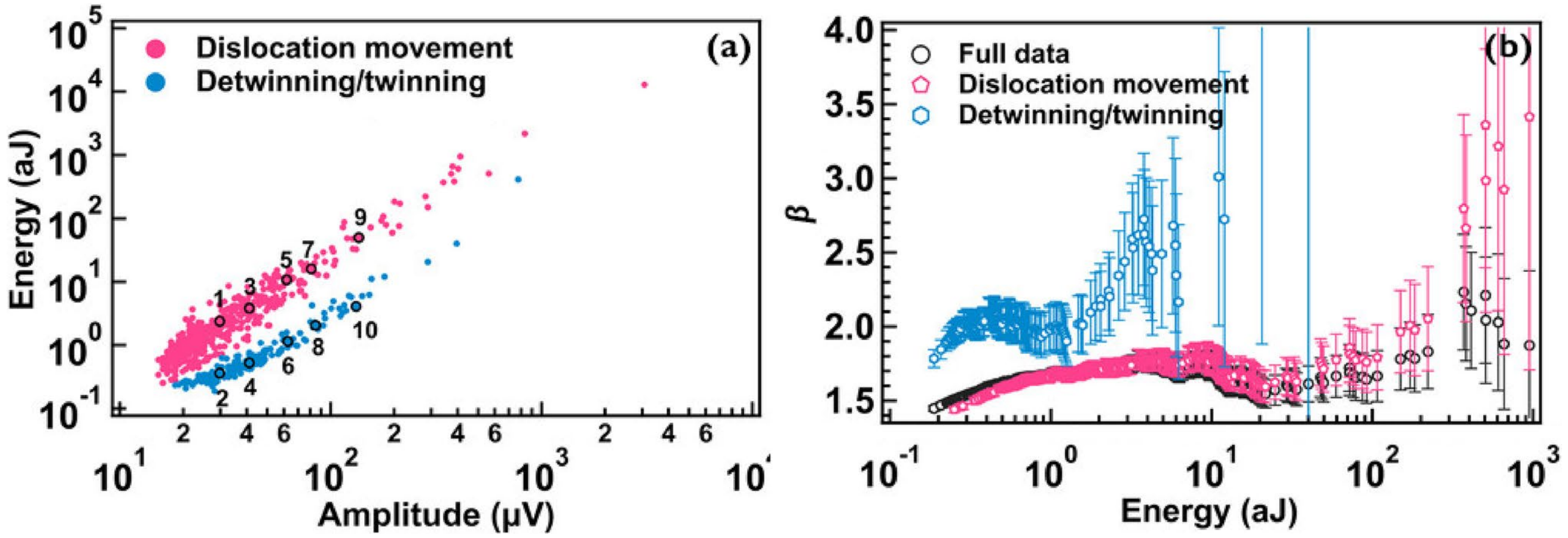

**Figure 7** Multiple avalanche populations in AE spectra of $Fe_{40}Mn_{40}Co_{10}Cr_{10}$ CCA. **a** Energy–amplitude relation showing two AE populations assigned to dislocation movement and detwinning/twinning. The two branches follow $E_i = \kappa_i A_i^2$, where $E_i$ is the AE energy of events in population $i$, $A_i$ is their peak AE amplitude and $\kappa_i$ is a branch-specific scaling prefactor reflecting differences in signal duration and waveform shape (the numbered points indicate representative AE events for which waveform examples were analyzed in the original work). The separation into distinct branches therefore indicates different avalanche mechanisms rather than one homogeneous event population. **b** Maximum-likelihood analysis of the AE energy distributions for the full dataset and for the two separated populations. Here, $\beta$ is the fitted exponent of the AE energy distribution, $g(E) \propto E^{-\beta}$, evaluated as a function of the lower energy cutoff. The dislocation-related and detwinning/twinning-related events show different scaling behavior, illustrating that AE intermittency in CCAs may contain multiple superposed event classes rather than a single avalanche population. (Adapted from [23] with permission).

"Atomistic simulations of intermittent plasticity in CCAs". These studies make clear that DSA is only one possible route to intermittency in CCAs, not a universal explanation. Related metastable CCA work on interstitially assisted TWIP/TRIP and twinning in CrMnFeCoNi-type systems is therefore relevant, even when it does not analyze avalanche statistics directly [104, 105]. These results help identify which deformation carrier should be expected before AE or serration statistics are interpreted. This mechanism dependence is summarized schematically in Fig. 8, which separates regimes where serrations are more likely to arise from solute-assisted dislocation pinning, cryogenic twinning or other low-temperature intermittent mechanisms.

## Atomistic simulations of intermittent plasticity in CCAs

Atomistic simulations provide a complementary view of intermittent plasticity by resolving nucleation, glide, pinning and structural rearrangements that are usually inferred indirectly from macroscopic experiments. They are particularly useful in CCAs, where local chemical environments, lattice distortion and SRO can be varied and their effects on dislocation motion examined directly. However, small simulation volumes, idealized initial defect structures and deformation rates far above conventional experiments require caution when quantitatively comparing simulated stress drops or avalanche exponents with bulk measurements [21, 103, 106].

The most direct atomistic connection to avalanche statistics was reported by Karimi et al., who simulated tensile deformation of FCC NiCoCrFeMn (Cantor alloy), NiCoCr and elemental Ni [103]. Following the initially high yield stress of the defect-poor simulation cells, plastic flow developed a stick-slip response with stress build-up intervals separated by rapid drops. At sufficiently low driving rates, the events showed robust power-law-like size and duration statistics, with avalanche-size exponents broadly consistent with the mean-field value of approximately 1.5 and with experimental ranges for crystalline plasticity. Increasing the deformation rate shifted the systems toward a subcritical regime with more rapidly decaying event statistics [103].

A particular advantage of atomistic analysis is that stress avalanches can be correlated directly with defect topology changes. In the Cantor alloy simulations, larger stress drops accompanied larger changes in locally HCP- and BCC-coordinated atomic populations, with the stronger correlation involving HCP stacking generated by partial dislocation slip [103]. Yet transformation-related structural changes did not produce a unique signature in the event-size distribution. Thus, even when an intermittent event can be linked directly to a structural rearrangement, its statistical exponent is not necessarily a unique mechanism label.

Other simulations address how chemical complexity controls the elementary barriers underlying collective behavior. Esfandiarpour et al. calculated dislocation depinning stresses in FCC VCoNi and examined the effects of lattice distortion and SRO on dislocation mobility [106]. For refractory BCC CCAs, simulated nanoindentation of HfNbTaTiZr and TaTiZrV, together with related HfNbTaZr calculations, shows that local chemical order can modify dislocation nucleation, trapping/detrapping, mobility and the spatial extent of the plastic zone [21, 107, 108]. These studies provide direct atomistic input for understanding how chemical complexity shifts the balance between dislocation–dislocation interactions and obstacle-controlled pinning.

Taken together, atomistic simulations establish causal links that are difficult to obtain experimentally: from local chemical environment to defect nucleation or depinning barriers, then to slip morphology and intermittent mechanical response. Their main limitation remains the scale transition, because molecular dynamics strain rates, finite dimensions, boundary conditions and simplified initial defect populations can strongly affect event statistics [21, 103, 106]. Atomistic simulations therefore provide a particularly valuable route for identifying elementary mechanisms and trends, while quantitative predictions of bulk avalanche exponents, PLC kinetics or macroscopic failure are best strengthened through validation across larger scales and broader experimental conditions.

## Serration amplitude as a failure-related metric

Besides identifying deformation mechanisms, serration statistics have recently been explored for predictive purposes. Hao et al. proposed the maximum stress drop magnitude during serrated tensile flow of a CoNiV MEA as a "failure threshold" [59]. In their experiments, serration amplitude increased with strain and fracture followed the largest serration, motivating the maximum drop as an operational indicator of proximity to failure. Here, the term denotes the largest serration expected during the tested deformation history rather than a conventional fracture-mechanics criterion such as critical stress-intensity factor, fracture strain or damage parameter.

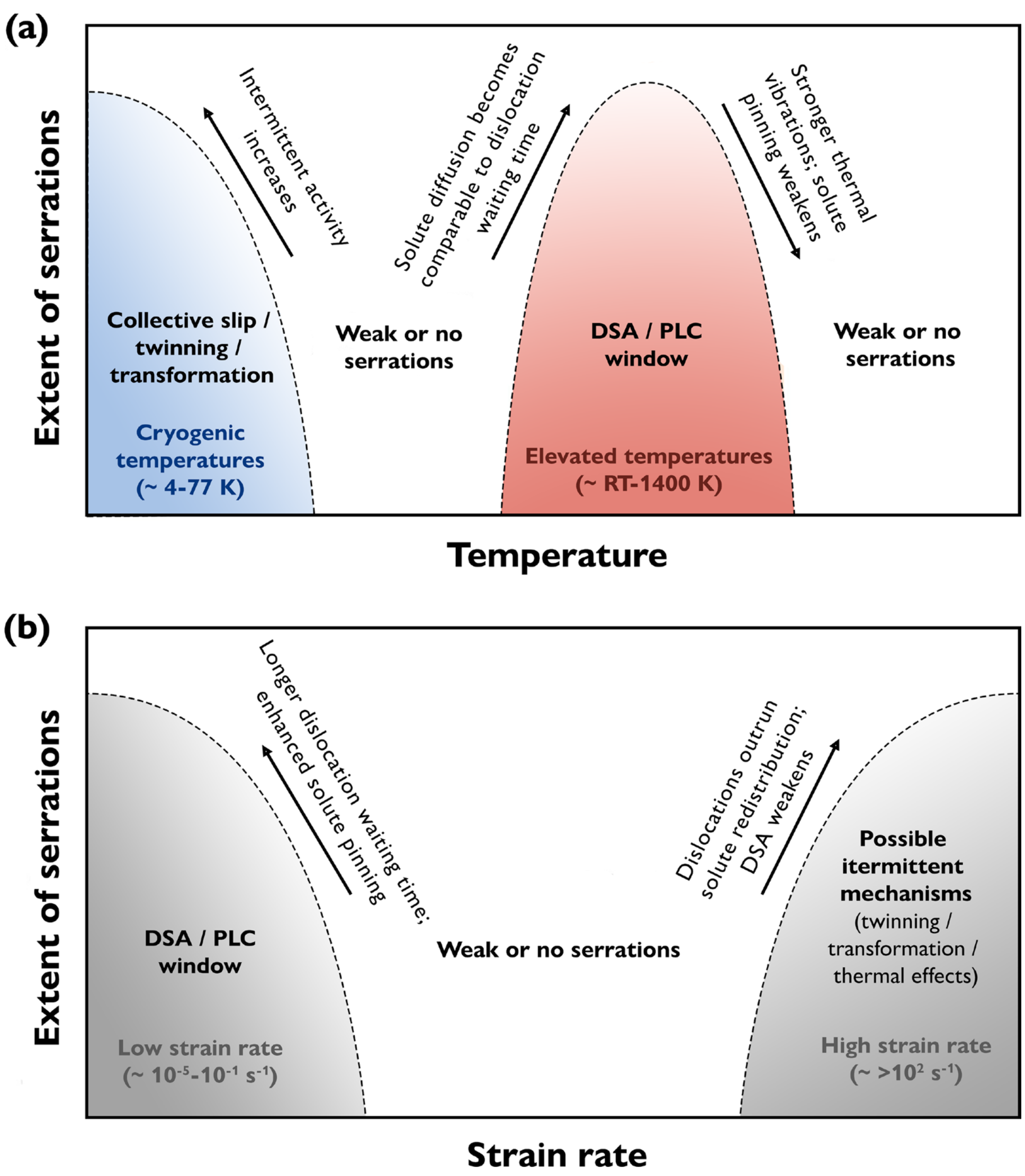

**Figure 8** Schematic temperature- and strain-rate-dependent regimes of serrated flow in CCAs. **a** At cryogenic temperatures, intermittent flow may arise from collective dislocation slip, twinning and/or martensitic transformation. With increasing temperature, DSA/PLC becomes possible when solute diffusion and dislocation waiting times become comparable, whereas at sufficiently high temperatures stable solute locking weakens and visible serrations may diminish or disappear. **b** At low to intermediate strain rates, sufficiently long dislocation waiting times can promote solute-assisted pinning and DSA/PLC-type serrations. Increasing strain rate may move the system out of the DSA window as dislocations increasingly outrun solute redistribution; at still higher rates, other intermittent mechanisms, including twinning, transformation and thermal effects, may become relevant. The boundaries between the indicated regimes are intentionally shown as diffuse because the transitions are gradual, may overlap, and depend strongly on alloy composition, crystal structure, phase stability, microstructure, temperature, strain rate and loading mode. The indicated ranges are therefore approximate, and the figure should not be interpreted as a universal mechanism map. (Adapted from [9]).

In the same study, the mean-field scaling of serration magnitudes was also combined with an in situ DSA pinning model and a contribution from deformation twinning [59]. The model reproduced the measured maximum stress drop amplitudes with errors of approximately 7–10% at the tested temperatures. The authors further combined the predicted maximum serration magnitude with a waiting time/pinning description to predict the engineering strain at which the maximum serration occurs. The approach therefore links an extreme event statistic to an evolving dislocation pinning mechanism rather than treating the largest serration as an empirical curve feature alone.

A related constitutive approach was developed for high-temperature tensile deformation of FCC CoCrFeMnNi [109]. Using lattice strain and dislocation density estimates together with the strain dependence of mobile and forest dislocation populations, the authors formulated a flow stress model with an explicit PLC contribution. The model links temperature-dependent serrated flow to the overall flow stress and thus connects microscopic dislocation populations, PLC instability and macroscopic constitutive response.

These studies illustrate a promising shift from descriptive serration analysis toward prediction, although the term "failure threshold" should still be interpreted cautiously [59]. The maximum serration need not coincide with fracture in other CCAs or under different loading conditions in the same alloy. Large stress drops may arise from DSA synchronization, twinning, transformation or other plastic instabilities without damage initiation, while fracture may occur without an exceptional preceding serration. The predictive value of such extreme event metrics should therefore be validated against independent damage indicators, fracture location, AE source populations, local strain fields and microstructural evidence before being treated as general failure criteria [9].

The evidence reviewed above shows why mechanism labels must accompany statistical labels. High-temperature serrations in many CCAs are most naturally discussed in the PLC/DSA framework, whereas cryogenic serrations in low stacking fault energy or metastable CCAs may reflect abrupt twinning, stacking fault activity, martensitic transformation or local thermal effects. Small-scale bursts may instead reflect source-controlled slip avalanches [9, 23, 24, 98]. Figure 8 should therefore be read as a qualitative map rather than a universal phase diagram: temperature and strain-rate shift the balance between competing deformation carriers, but the exact boundaries depend on composition, phase constitution, stacking fault energy, grain size, defect density and testing geometry. These intermittent responses need not belong to the same physical class.

## Mechanism- and structure-specific synthesis

The occurrence or morphology of a stress serration is not sufficient to identify its microscopic origin [9, 22]. Mechanism attribution should instead combine several levels of evidence: the relevant temperature–strain-rate window, direct or indirect identification of the active deformation carrier, the spatial organization of strain, and temporal correlation among mechanical, AE, imaging and structural signals [7, 12]. The principal mechanism classes relevant to CCAs are summarized in Table 2.

DSA and the PLC effect provide the most clearly identifiable route to macroscopic serrations when the waiting time of temporarily arrested dislocations becomes comparable to the redistribution time of mobile solutes [12, 74, 75]. The strongest evidence is therefore not simply designation of a curve as type A, B or C, but the combined observation of a restricted temperature–strain-rate window, negative strain-rate sensitivity, systematic response to strain rate changes and, where available, propagating or intermittently nucleating deformation bands [7, 13, 22]. Stress drop/AE correlation supports the occurrence of rapid collective releases, but does not by itself identify the diffusing species or establish classical solute aging.

A second class comprises jamming- or interaction-dominated collective dislocation dynamics. Here, barriers to motion are generated primarily by the dislocation population itself through long-range anisotropic interactions, junction formation and evolving metastable configurations. Such systems may exhibit strain bursts, AE activity and broad event statistics even without a fixed population of strong pinning obstacles [60, 61]. The associated critical-like behavior may extend over a range of stresses and should not automatically be interpreted as conventional depinning.

A distinct limit is reached when relatively persistent microstructural obstacles dominate the resistance to dislocation motion. Solute fields, precipitates, interfaces, strong local chemical environments or other forms of quenched or slowly evolving disorder may then generate a pinning landscape from which dislocations or correlated dislocation structures escape collectively. Simulations show that increasing obstacle strength can shift avalanche behavior from a jamming-dominated regime toward depinning-like criticality [62, 63]. In CCAs, local chemical environments and SRO can strongly modify dislocation nucleation and mobility, providing a possible microscopic route by

**Table 2** Mechanism-based interpretation of intermittent and serrated plasticity in CCAs

| Mechanism or physical regime | Favored structural or phase context | Physical basis and conditions | Most diagnostic evidence | Main interpretive caution | Representative references |
|---|---|---|---|---|---|
| DSA/PLC instability | Reported in both FCC and BCC CCAs, commonly within intermediate or elevated-temperature windows; interstitial additions may shift the operative range | Time-dependent pinning of dislocations when dislocation waiting and solute redistribution times become comparable | Restricted temperature–strain-rate window; negative strain-rate sensitivity; strain-rate jump response; DIC or extensometry evidence of PLC bands; correlated AE and stress drops | Serration type alone does not establish DSA. Time-dependent solute pinning should also not be conflated automatically with the quenched disorder depinning problem used in statistical-mechanical avalanche models | [7, 9, 12, 13, 22] |
| Jamming- or interaction-dominated collective slip | Relatively weak external obstacle landscape; collective dislocation activity in FCC or BCC systems, especially during microplasticity and small-scale deformation | Mutual long-range dislocation interactions generate evolving metastable structures and internal barriers; arrest is predominantly self-generated rather than imposed by fixed obstacles | Strain/displacement bursts; dislocation structures; slip band evolution; temporal correlations; stress-dependent avalanche statistics | Does not require quenched disorder and is not equivalent to classical depinning. Broad or power-law-like statistics may occur over an extended stress range and need not imply a unique critical point | [60, 61, 63] |
| Obstacle-pinning/depinning-dominated collective slip | Systems containing sufficiently strong persistent disorder, including precipitates, interfaces or strong spatial variations in local resistance | Collective escape of dislocations from an externally or microstructurally imposed pinning landscape; sufficiently strong obstacle interactions may generate a depinning-like critical regime | Dependence of event statistics and cutoffs on obstacle density or strength; source/obstacle-resolved imaging or simulation; systematic evolution toward a critical driving condition | Power-law-like statistics alone do not prove depinning. The crossover from jamming to pinning depends on the relative strengths of obstacle and dislocation–dislocation interactions | [21, 62, 63, 106] |
| Deformation twinning | Particularly relevant to low stacking fault energy and metastable FCC CCAs, often under cryogenic or high stress conditions | Abrupt twin nucleation, propagation or thickening | TEM/EBSD identification; crystallographic orientation dependence; diffraction evidence; mechanism-separated AE populations | Twinning frequently coexists with dislocation slip; a distinct event population should be demonstrated before assigning statistics specifically to twinning | [20, 23, 105] |

**Table 2** continued

| Mechanism or physical regime | Favored structural or phase context | Physical basis and conditions | Most diagnostic evidence | Main interpretive caution | Representative references |
|---|---|---|---|---|---|
| Martensitic or transformation-mediated intermittency | Metastable FCC CCAs undergoing FCC to HCP transformation and high-entropy shape-memory systems | Nucleation and propagation of a stress- or strain-induced transformation front | In situ diffraction or phase fraction evolution; transformed microstructure; Lüders-like band propagation; transformation-related AE; thermal recovery | Transformation bands and stress serrations may resemble PLC behavior; mechanical curve morphology alone cannot distinguish them | [20, 24, 104] |
| Precipitate-, phase boundary- or interface-modulated intermittency | Multiphase, ordered, dendritic or precipitation-strengthened CCAs with FCC or BCC matrices | Interfaces and second phases modify local stresses, source operation, mean free path and event arrest or synchronization | Controlled comparisons among single phase, multiphase or differently aged states; phase-resolved imaging; changes in localization, event cutoff or burst rate | Usually modifies another carrier—slip, twinning or transformation—rather than constituting a unique avalanche mechanism | [9, 18, 56, 110] |
| Damage- or thermal instability-related events | Potentially relevant near failure, at cryogenic temperature, high rate or under strongly localized deformation | Microcracking, decohesion, cavitation or localized thermal softening releases strain or elastic energy abruptly | In situ or post-mortem damage imaging; stiffness degradation; fracture surface evidence; temperature measurements; separation from deformation-related AE populations | These sources should be excluded before large events are assigned to dislocation avalanches, DSA, twinning or transformation | [2, 9, 98] |

which chemical complexity changes this competition [21, 27, 28, 106]. Nevertheless, identifying a depinning regime requires more evidence than a power-law-like event distribution alone.

Twinning and martensitic transformation represent distinct carriers of intermittent strain. Twin nucleation and growth, or nucleation and propagation of a transformation front, may produce AE bursts, Lüders-like bands and macroscopic serrations [23, 24, 100]. These mechanisms should be assigned only when supported by microscopy, diffraction, crystallographic orientation relationships or phase fraction evolution [20, 104, 105], or by mechanism-resolved event populations [23, 24]. Similarity to a PLC-type mechanical curve is insufficient because transformation- or twinning-mediated localization can produce a superficially comparable response.

Precipitates, phase boundaries, dendritic segregation and local chemical order generally modify rather than uniquely define the deformation carrier [9, 18]. They may arrest, fragment or synchronize dislocation motion, alter the upper event-size cutoff, or promote twinning and transformation by modifying local stresses and fault energy landscapes [20, 21, 56]. Their role is most convincingly established through controlled comparisons of related microstructural states. Damage, cracking and local thermal softening should also be considered as competing sources of abrupt signals, particularly when large events occur close to failure [9].

For CCAs, the nonequilibrium-thermodynamic viewpoint provides a useful way of organizing the consequences of chemical and microstructural heterogeneity without treating chemical complexity itself as a deformation mechanism. Spatial variations in lattice resistance, local fault energies, SRO, precipitates and phase boundaries modify both the distribution of local reaction barriers and the internal stresses generated during deformation [21, 103]. Within the framework of McDowell and Liu, such heterogeneity may be viewed as modifying the probabilities of local dislocation-barrier reactions and the evolution of their degree of correlation [64, 65]. This suggests a physically testable question for CCAs: whether changes in composition, SRO or phase constitution primarily broaden and decorrelate the population of elementary events, or instead promote sufficiently correlated activation that local activity survives coarse graining and becomes visible as specimen-scale intermittency. At present, conventional stress drop, AE or DIC measurements do not provide a direct experimental measurement of the thermodynamic degree-of-correlation, and connecting this quantity quantitatively to CCA experiments will require dedicated simulation-experiment comparisons.

Table 2 is intended as a diagnostic guide rather than an exhaustive catalog of testing conditions. The principal distinction is between the measured signal, the active carrier of irreversible strain, and the microstructural or kinetic factor controlling that carrier. DSA, collective slip, twinning and martensitic transformation are possible physical origins of intermittent strain; AE bursts, stress drops, DIC bands and pop-ins are experimental manifestations; and chemical disorder, SRO, precipitates and phase boundaries are control variables that can modify several mechanisms [2, 9].

The most convincing mechanism assignments are therefore those supported by independent evidence. FeAlCrMo combines negative strain-rate sensitivity, systematic serration changes, strain-rate jump behavior and AE-stress drop correlations, supporting a DSA/PLC interpretation [7, 13]. In contrast, studies of FeCoNiTiAl and $Al_{0.3}CoCrFeNi$ demonstrate that spatial and temporal intermittency may remain distributed below the threshold for visible macroscopic serration [5, 25]. Mechanism-resolved AE and structural characterization in $Fe_{40}Mn_{40}Co_{10}Cr_{10}$ and high-entropy shape-memory alloys further show that twinning or martensitic transformation may generate event populations distinct from ordinary dislocation slip [23, 24]. These examples illustrate why signal classification should follow, rather than replace, mechanism identification.

Small-scale tests and atomistic simulations provide complementary access to source activation, local slip band development and defect–obstacle interactions [6, 8, 102, 103]. They are particularly useful for testing how chemical disorder and SRO alter the earliest irreversible events [21, 106], but direct quantitative transfer to bulk CCA deformation remains limited by specimen size, free surfaces, simulation strain rate, initial defect structure and instrument response [2, 4].

### *Crystal structure and phase constitution perspective*

FCC CCAs currently provide the largest body of evidence reviewed here. Their intermittent response is strongly conditioned by stacking fault energy and phase stability. Relatively stable FCC alloys may exhibit full dislocation slip and elevated-temperature DSA, whereas lower stacking fault energy promotes partial dislocation activity and deformation twinning. In metastable systems, faulting may develop further into FCC-to-HCP martensitic transformation [12, 18, 20]. Stress serrations or AE bursts in FCC CCAs therefore cannot be interpreted without distinguishing among full dislocation slip, partial slip, twinning and transformation [23].

BCC CCAs provide a different physical limit because deformation is additionally conditioned by comparatively high lattice friction and screw dislocation mobility, producing strong temperature, rate and size dependence [8, 18]. The FeAlCrMo studies nevertheless show that a BCC CCA may exhibit a well-defined DSA/PLC window supported by strain-rate sensitivity, rate jump testing and AE [7, 13]. Refractory BCC CCA micropillar and atomistic studies probe source activation and SRO effects at much smaller scales [8, 21, 107]. Direct comparisons should consequently separate intrinsic BCC lattice resistance from time-dependent solute pinning and external size effects.

Within the literature surveyed here, direct avalanche-resolved evidence for HCP CCAs remains limited. Conventional HCP crystals remain important reference systems because restricted slip and comparatively weak screening can produce well-separated AE avalanches [3, 11], but these results should not be transferred automatically to chemically complex HCP alloys. Future HCP CCA studies should distinguish the contributions of basal, prismatic and pyramidal slip, twinning, texture and phase constitution before comparing their apparent scaling behavior with that of simple HCP crystals [18]. The scarcity of such data should be presented as an evidence gap rather than evidence for common CCA behavior.

Multiphase and metastable CCAs cut across the FCC, BCC and HCP categories. Phase boundaries, ordered precipitates, dendritic segregation and transformation fronts generate additional internal length scales, local stress concentrations and load-partitioning effects [9, 18]. These features may suppress system-spanning slip by interrupting dislocation motion or, conversely, promote large synchronized events through interfacial stress accumulation and transformation [20, 56]. In such materials, phase constitution and phase stability may be more direct controls of intermittency than the CCA designation itself.

Taken together, the reviewed evidence does not support “CCA” as an independent deformation mechanism or as a sufficient predictor of avalanche behavior. Crystal structure, stacking fault energy, phase stability, microstructure, temperature, strain rate and loading geometry provide the more direct physical control variables, while chemical complexity modifies their local energetic and kinetic landscapes [9, 18, 21].

## Comparison of reported power-law exponents

Power-law exponents are useful only when the measured quantity and analysis method are specified. AE energy, squared AE amplitude, stress drop magnitude, local strain-rate burst size and nanopillar displacement burst size are not the same physical observable. In Table 3, the tabulated exponents are reported, wherever possible, as positive PDF or PDF-equivalent exponents, $p(x) \propto x^{-\beta}$. This convention matters because a CCDF scales as $P(>x) \propto x^{-(\beta-1)}$ for the same underlying power law, i.e., with a slope milder by one in absolute value. When a cited paper plots a CCDF, Table 3 therefore uses the authors’ stated distribution exponent or a PDF-equivalent exponent, not the raw CCDF slope, unless explicitly noted. PDFs, CCDFs and binned histograms may otherwise yield different apparent numerical slopes unless they are converted consistently. Table 3 should consequently be read as a comparison of reported regimes and observables, not as a strict ranking of materials. These comparisons should therefore be regarded as qualitative rather than strictly quantitative, because the reported alloys differ not only in composition but also in processing route, thermo-mechanical history, grain size, phase constitution, defect density and testing conditions. Because the reviewed studies differ substantially in event definition, specimen size and fitting procedure, the term "power-law-like" is used where a heavy-tailed or approximately linear log-log regime is reported without a formal goodness-of-fit test and comparison with alternative distributions [57, 58].

A tentative but useful pattern nevertheless emerges. HCP crystals such as ice, zinc and cadmium often show AE energy exponents close to 1.5, consistent with wild, scale-free avalanche dynamics [3, 10, 11, 66]. In micron-scale Zn single crystals, coupled AE/micropillar experiments give closely related but not identical exponents for stress drops

**Table 3** A non-exhaustive list of approximate experimental power-law exponents reported for selected material classes and observables. Exponent $\beta$ denotes a positive PDF or PDF-equivalent exponent, $p(x) \propto x^{-\beta}$. The values are not strictly interchangeable because the event definitions, probability representations and measured quantities may differ. The tabulated values summarize the published analyses and should not be interpreted as an independent reanalysis of the original event datasets

| Material class/system | Observable/event definition and distribution | Representative $\beta$ exponent(s) | Interpretation/caution | Refs. |
|---|---|---|---|---|
| HCP single crystals (ice, Zn, Cd) | AE energy PDF, $p(E) \sim E^{-\beta}$, Cd/Zn-Al compared slip- and twinning-dominated stages. | Close to ~1.5 in all cases. | Wild plasticity; strong anisotropy and weak screening of long-range interactions. Slip and twinning gave different waveforms but indistinguishable AE energy PDFs in Cd and Zn. Possible coexistence of mild background and power-law wild bursts in ice | [3, 10, 11, 66] |
| Micron-scale Zn single crystals | Coupled stress drops and AE events during Zn micropillar compression; PDFs for stress/force drops and AE energy | Earthquake-like scaling; both stress drop and AE energy exponent close to 5/3 and Omori-type aftershock behavior | Clean bridge between slip avalanches and AE: basal slip only in the analyzed pillars, with twinning excluded | [50] |
| Bulk FCC examples (Cu, Al) | AE burst energy and high-resolution plastic fluctuations, mainly during early deformation or cyclic loading | FCC Cu shows scale-free bursts in early deformation but with low event density; cyclic Al gives AE energy exponent of ~1.7-1.9 | Avalanches in FCC more sporadic than in HCP crystals. Strain hardening and dislocation substructure shift the response toward mild plasticity | [3, 66, 111–113] |
| Conventional PLC alloys | Stress drops and AE intensity | Type-A stress drop examples near ~1.25; AE event energies in Al-Mg-type alloys ~2.5-3.6 | Stress serrations and AE need not have the same statistics because stress drops are macroscopic synchronized events, whereas AE resolves faster internal rearrangements | [52, 74, 78] |
| Small FCC/BCC crystals, micropillars | Displacement/slip bursts and stress-integrated avalanche sizes; exact observable depends on experiment | Often ~1.5-1.7 in mean-field-like regimes; broader values occur with orientation, drive rate, size and hardening | Exponent varies with size, disorder and hardening. Smaller is often wilder, but nontrivial exponents should not be read as one universal value | [3, 51, 67–72] |
| CCA nanopillars | Slip-size/displacement-jump avalanches | Power-law or truncated power-law behavior; exponent ~1.5 | Direct imaging links bursts to dislocation avalanches, bands and slip-line formation | [6, 97, 102] |
| FeAlCrMo CCA | Squared AE amplitude, $p(A^2) \sim (A^2)^{-\beta}$ | Reported $\beta \sim 1.4$ (300 °C plastic), ~1.6 (400 °C plastic), ~1.7 (microplastic) | CCA evidence for power-law AE during microplasticity and DSA/PLC | [7] |
| $Al_{0.3}CoCrFeNi$ CCA | AE events in macroscopically smooth flow; normalized AE intensity/squared-amplitude-type statistics | Reported range ~1.5-2.1 depending on condition and signal regime | Mesoscopic complexity; not directly comparable without event-definition details | [5] |

**Table 3** continued

| Material class/system | Observable/event definition and distribution | Representative $\beta$ exponent(s) | Interpretation/caution | Refs. |
|---|---|---|---|---|
| $Fe_{40}Mn_{40}Co_{10}Cr_{10}$ CCA | Branch-resolved AE energy populations separated by energy-amplitude scaling | Two distinct AE branches rather than one exponent. $\beta \sim 1.7$ for dislocation-related AE; $\beta \sim 2.0$ for detwinning/twinning-related AE | Mechanism-resolved AE is essential when slip and twinning coexist; one dataset may contain superposed event classes | [23] |
| Shape-memory $Cr_{20}Mn_{20}Fe_{20}Co_{34.4}Ni_{5.6}$ CCA | AE during Lüders-band-type martensitic transformation and thermal recovery; PDFs of energy $E$, amplitude $A$ and duration $S$ | Exponents comparable to some shape-memory systems within error bars. Transformation: $\beta_E \sim 1.8$, $\beta_A \sim 2.7$, $\beta_S \sim 2.2$; recovery: $\beta_E \sim 1.8 \pm 0.1$, $\beta_A \sim 2.7$, $\beta_S \sim 2.1$ | Transformation avalanches should be compared with slip avalanches only with clear mechanism labels | [24] |
| CCA atomistic simulations | Event definition: Stress drops/slip-avalanche sizes in NiCoCr and NiCoCrFeMn atomistic simulations | Scale-free behavior in quasi-static regimes; avalanche-size scaling close to ~ 1.5 under suitable conditions | Useful for hypotheses, but simulation strain rate, size and potential choice limit direct comparison | [103] |
| Al and Al-alloy micropillars with tunable disorder | Displacement bursts in micropillar compression; fraction of plasticity released by wild fluctuations, $W$ | $\beta \sim 1.5–1.8$ for small/wild pure Al pillars; larger apparent $\beta$ in crossover regimes with stronger disorder; Gaussian/mild cases do not give a robust exponent | Shows "smaller is wilder" and "dirtier is milder;" useful conceptual bridge for CCA disorder and precipitate effects | [73] |

and AE energy, with $\beta \sim 1.8$ for stress drop sizes and $\beta \sim 1.7$ for AE energies [50]. This illustrates that different observables sample different projections of the same intermittent process. Micron-scale crystals and nanopillars can show related intermittent slip, but their exponents and cutoffs depend on size, crystal orientation, drive rate and hardening.

For instance, Zhang et al. demonstrated this explicitly in compressed Al and Al alloy micropillars with diameters from about 500 to 6000 nm [73]. In pure Al, where the effective pinning strength was only about 4.5 MPa and the corresponding internal length scale was about 1537 nm, small pillars remained strongly intermittent and the burst size exponent approached the mean field-like value of 1.5 for the smallest sizes, ~1.6–1.9 for 500–2000 nm pillars, and ~2.4 for 6000 nm pillars. Introducing quenched disorder progressively moved the same size range toward milder fluctuations. From Al–Sc clusters to Al–Sc precipitates and Al–Cu–Sn, the pinning strength increased from about 28 to 71 MPa, the internal length scale decreased from about 251 to 98 nm, and the burst statistics evolved from power-law-like behavior with larger exponent toward almost Gaussian fluctuations. Thus, alloying and quenched disorder do not simply change the strength; they introduce internal pinning length scales that compete with the external pillar size and can shift plasticity from wild, power-law-dominated fluctuations toward milder, more Gaussian-like flow [73].

Bulk FCC metals occupy an intermediate position between wild HCP plasticity and the more strongly averaged macroscopic response of work-hardening metals. In Cu, AE and high-resolution extensometry show scale-free but sparse plastic fluctuations mainly during early deformation, before dislocation substructures reduce the avalanche activity [111]. Cycled Al illustrates the same transition under repeated loading: AE-burst-energy distributions during the initial hardening stage give PDF-equivalent exponents of ~1.7–1.9, while the AE wildness and burst rate decrease rapidly as dislocation patterns form [3, 112]. Cyclic Cu measurements connect this loss of wildness to the formation, destruction and rebuilding of dislocation structures and the associated change in dislocation mean free path [113]. Thus, bulk FCC plasticity is generally milder than HCP plasticity not because avalanches are absent, but because multislip, forest hardening and evolving dislocation substructures screen long-range correlations more effectively [3, 111–113].

Conventional PLC alloys illustrate a different comparison problem. Some studies analyze macroscopic stress drop amplitudes or durations, whereas others analyze AE intensity, AE energy, or local strain rate bursts measured by local extensometry [52, 74, 78]. In AlMg alloys, type-A stress serrations and low-amplitude stress drops may show $\beta$ exponents in the range 1–1.5. By contrast, AE statistics in PLC alloys are often steeper: squared AE amplitude, $A^2$, distributions may give $\beta \sim 2.5$–2.9 in Al-Mg, $\beta \sim 3.6$ before and ~2.4 after PLC onset in AlMgScZr, and more generally values from about 2 to 3 in binary AlMg alloys and sometimes up to 4 in precipitate-containing AlMg-based alloys. This difference is not surprising: stress drops are affected by band propagation, synchronization of many local events, and machine-sample coupling, whereas AE records elastic waves emitted by rapid internal rearrangements. Indeed, AE amplitudes during stress drops may remain in the same range as during smooth reloading, while the stress drops are associated mainly with clustering or merging of many AE hits, producing longer apparent AE durations. Their exponents may therefore differ even when both signals originate from the same unstable DSA/PLC process [78].

The power-law-like exponents reported for FeAlCrMo are interesting because the same material shows both microplastic AE and macroscopic DSA/PLC serrations [7]. The reported values, roughly 1.4 and 1.6 at 300 and 400 °C, respectively, overlap with simple crystal avalanche statistics, but this overlap should not be taken as proof of identical mechanisms. It indicates that the measured AE event hierarchy has a similar broadness. The comparison should therefore be used to define hypotheses. If CCA chemical disorder mainly acts as quenched pinning, one may expect fewer system-spanning events and a steeper apparent distribution. If DSA synchronizes many elementary avalanches into macroscopic drops, the same alloy may show large stress serrations but AE exponents that still resemble those of conventional avalanche plasticity. FeAlCrMo illustrates this ambiguity: the stress curve shows apparently synchronized PLC events, but the AE-derived $A^2$ exponents remain close to values reported for simpler crystalline systems [3, 7, 11].

Another useful comparison is between microplastic and plastic exponents within a single alloy. In the same FeAlCrMo system, the microplastic value of $\beta$

~ 1.7 is slightly higher than the plastic flow values at 300 and 400 °C [7]. A higher exponent corresponds to a relatively smaller contribution from the largest events. One possible interpretation is that early mobile source activation is more spatially distributed, whereas DSA during plastic flow can synchronize larger dislocation groups. This remains a hypothesis until simultaneous spatial information or mechanism-resolved AE is available.

The main implication is that, although the number of direct CCA studies reporting avalanche exponents is still very small, the available CCA exponents currently overlap with those of simpler systems. This overlap undermines the claim of a CCA-specific universality class. The more specific CCA contribution is probably different: several event populations may coexist and sometimes become synchronized. Microplastic dislocation avalanches, DSA-related unpinning, twinning bursts, transformation bursts and damage-related events can all occur within the same broad material family.

Table 3 is deliberately conservative—the purpose is not to reduce all experiments to a single exponent, but rather to show that the reported values occupy overlapping ranges while being measured through different experimental channels. A stress drop exponent mainly reflects the macroscopic instability and boundary conditions. An AE energy exponent reflects how rapidly elastic energy is radiated by internal rearrangements. A micropillar displacement burst exponent reflects discrete plastic strain increments under small volume boundary conditions. The same material could therefore reasonably give different numbers in different columns.

Moreover, Weiss et al. show that mild and wild fluctuations can coexist in a single material class, while Zhang et al. show that the wild fraction can be tuned by the competition between external sample size and internal disorder length scale [66, 73]. This is directly relevant to CCAs because chemical disorder, SRO, precipitates and phase boundaries provide several internal length and pinning scales that may either fragment avalanches or synchronize them into larger events.

The most informative future comparisons will be internal rather than purely cross-material. The same alloy should be tested before and after aging, after prestraining, or across a controlled temperature/strain rate series, while keeping the AE settings, DIC resolution and fitting procedure fixed. Such paired comparisons can show whether SRO, precipitates, phase stability or prior defect density changes the exponent, the upper cutoff, or only the event rate. This approach is more informative than comparing isolated exponent values from unrelated materials.

## Synthesis: what CCA intermittency adds to avalanche plasticity

The studies reviewed above suggest that intermittency in CCAs should be treated as a scale-dependent deformation problem rather than as a single phenomenon. At the microscopic scale, intermittent plasticity reflects the activation, arrest and interaction of dislocation sources and local defect configurations. At the mesoscopic scale, these events may remain spatially distributed, or they may organize into localized slip bands, transformation fronts or deformation bands. At the macroscopic scale, only a subset of this activity becomes visible as stress serrations. The central question is therefore not simply whether a given CCA shows serrated flow, but how chemical complexity, SRO, metastability and microstructure alter the coupling between elementary events and specimen-scale localization.

### Established trends

Several conclusions appear robust. First, intermittent plasticity is not restricted to macroscopically serrated flow. AE, local strain measurements and micromechanical tests show that discrete defect activity can occur during nominally smooth deformation and below conventional yielding [4, 5, 7, 10, 11, 25, 49, 50, 52, 114]. Second, DSA/PLC remains one important route to macroscopic serrations in CCAs, especially at elevated temperatures and strain rates where solute–dislocation timescales become comparable [7, 9, 12, 13, 22, 74, 75, 89–91, 93–95]. Third, CCA-specific features such as local chemical fluctuations, SRO, low stacking fault energy, metastability and multiphase microstructures can modify how plastic events are arrested, redistributed or synchronized [20, 21, 23, 24, 26–31, 54, 98–101, 104, 105]. These features do not replace the general physics of dislocation avalanches, but in CCAs they can coexist and be tuned over an unusually broad compositional and microstructural range.

This view also explains why apparently contradictory observations can coexist. Smooth macroscopic flow does not imply homogeneous deformation, because AE, DIC and fine stress-fluctuation analyses may reveal hidden intermittency below the load cell resolution [5, 25]. Conversely, strong macroscopic serrations do not necessarily represent elementary avalanches, because a single stress drop may be the synchronized outcome of many smaller dislocation events or local structural rearrangements [7, 13, 22, 74, 78, 93–95]. The useful distinction is therefore not simply between serrating and non-serrating alloys, but between distributed, weakly synchronized and strongly synchronized plastic activity.

## Open problems and possible overinterpretations

At the same time, several interpretations should remain cautious. Power-law-like distributions alone do not prove self-organized criticality, nor do similar exponents necessarily imply a common microscopic mechanism. Stress drop statistics, AE energy statistics and slip event statistics probe different observables, different time scales and different degrees of coarse graining. Their exponents may therefore differ even when the underlying deformation process is related [4, 7, 52, 55, 56, 74, 78]. Similarly, the presence of serrations does not uniquely identify DSA, because twinning, martensitic transformation, cracking, adiabatic heating or collective slip localization can produce superficially similar mechanical signatures [9, 23, 24, 98–102]. For CCAs, this ambiguity is amplified by the coexistence of chemical disorder, phase heterogeneity and multiple competing deformation modes.

The most important unresolved issue is causality. It is plausible that local chemistry, SRO, fluctuating fault energies, dendritic segregation, precipitates and phase boundaries affect avalanche nucleation and arrest, but direct links between a measured microstructural state and a particular avalanche exponent remain scarce. A change in exponent, cutoff or event rate may reflect altered dislocation mobility, source density, hardening, detection threshold, localization length scale or a change in deformation carrier. For this reason, statistical analysis should be treated as a complement to mechanism identification, not as a replacement for it.

Adjacent work remains useful mainly as boundary conditions for this interpretation. Nonlinear and dynamical system treatments of the PLC effect define the background for band dynamics beyond simple curve classification [115]. CoNiV serration statistics and atomistic simulations extend the concentrated alloy space in which avalanche-like plasticity can be tested [106, 116]. Small-scale CrMnFeCoNi deformation and metastable TWIP/TRIP alloy design studies define complementary limits where sample size, stacking fault energy, twinning and transformation may dominate over classical DSA [18, 117–119].

## From local intermittent dynamics to continuum response

The coexistence of strong local heterogeneity with a smooth macroscopic stress-strain curve is not contradictory. A conventional continuum response represents spatial and temporal coarse graining over many elementary defect processes. When the representative volume and observation time contain a sufficiently large number of weakly correlated events, their individual fluctuations are averaged and the bulk response becomes effectively smooth. Conversely, when the correlation length, event size or synchronization time becomes comparable with the relevant specimen or observation scale, individual events or clusters remain experimentally visible as strain bursts, AE activity, localized bands or stress serrations [3, 4, 120].

The constrained local equilibrium construction of McDowell and Liu provides one formal route for this scale transition [64]. Grains, phases or other subsystems are assigned evolving local thermodynamic states, while the macroscopic constitutive response emerges from the ensemble of these heterogeneous subsystems and their irreversible transitions. Internal stresses and defect populations can consequently enter reduced-order constitutive models as internal state variables rather than being represented only through a spatially homogeneous flow stress. Related coarse-grained dislocation theories likewise demonstrate that continuum internal stress and back stress terms can emerge from explicitly heterogeneous dislocation structures [121].

For avalanche plasticity, the important implication is that averaging removes information as well as noise. A smooth macroscopic constitutive law may reproduce mean flow stress and hardening while remaining insensitive to the distribution, spatial organization and temporal correlation of the underlying events. The experimental approaches reviewed here therefore provide information complementary to conventional constitutive data: AE emphasizes rapid internal rearrangements, DIC resolves spatial localization, and small-scale mechanical testing reduces the averaging volume. A predictive multiscale description should ultimately reproduce both the continuum-level constitutive response and the statistics and spatial organization of the intermittent processes from which that response emerges [2, 64, 121].

Table 1 distinguishes the experimental observables, whereas Table 2 summarizes the principal physical mechanisms and the evidence required for their attribution. Figure 9 integrates these two levels into a general sequence for interpreting event statistics [2, 9]. For dislocation-mediated events, the jamming–pinning/depinning distinction and the degree of correlation provide an additional interpretive level between the microscopic carrier and the measured statistics [63, 65].

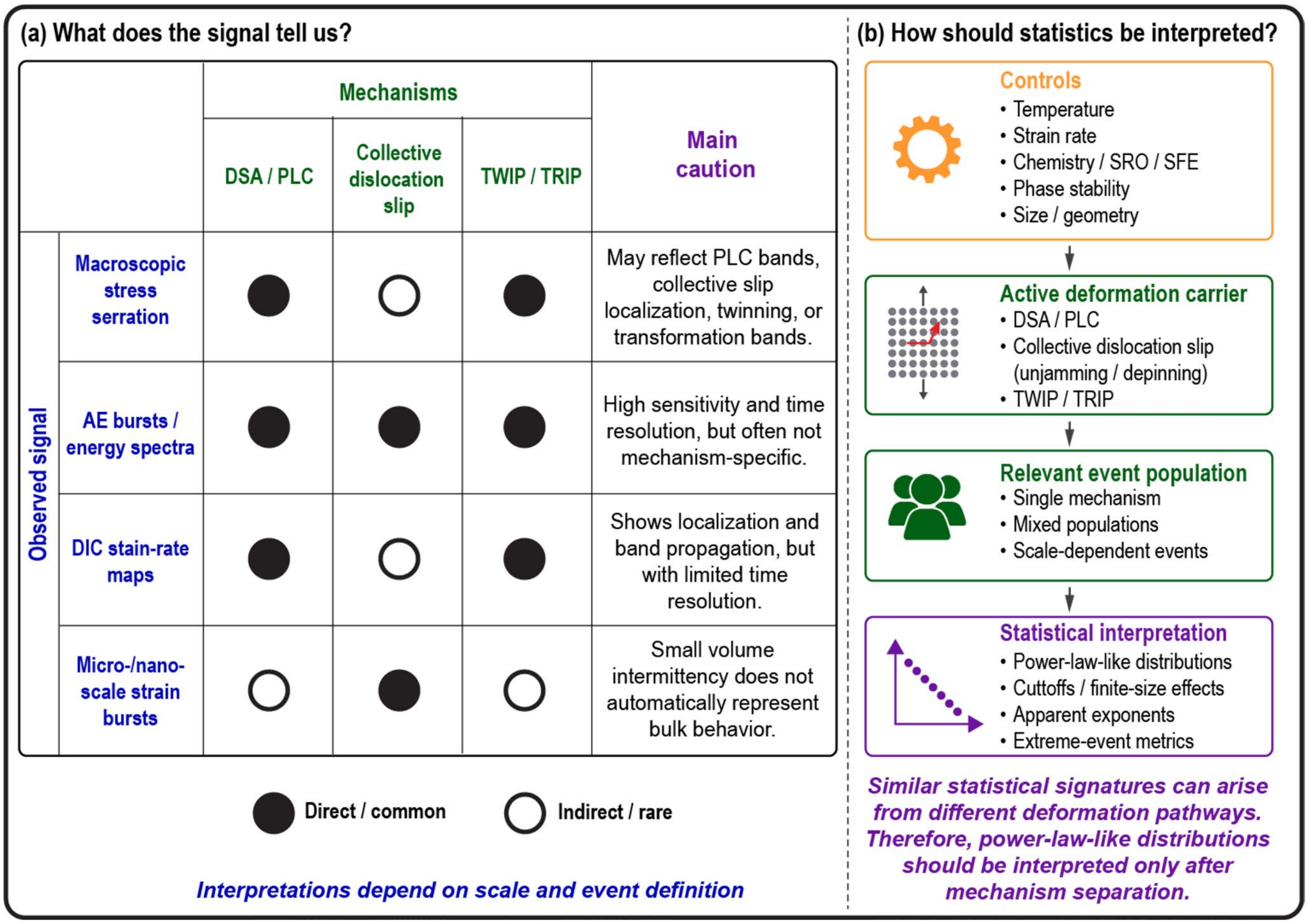

**Figure 9** Framework for interpreting intermittent plasticity in CCAs. **a** Experimental observables are not unique mechanism labels: macroscopic stress serrations, AE bursts or energy spectra, DIC strain-rate maps and micro-/nanoscale strain bursts may originate from DSA/PLC, collective dislocation slip, twinning or martensitic transformation [2, 9, 23]. Filled symbols indicate common or direct associations, whereas open symbols indicate indirect or less frequent associations. The correspondence between signal and mechanism is therefore scale- and event definition-dependent, and complementary structural or spatial information is generally required for reliable mechanism identification [2, 5, 7, 23]. **b** Statistical interpretation should proceed from the testing conditions and microstructure, through identification of the active deformation carrier and the relevant event population, to analysis of distributions, cutoffs, apparent exponents and extreme event metrics [3, 56, 59]. For collective dislocation slip, the underlying interaction regime may range from predominantly dislocation–dislocation interaction-controlled jamming to obstacle-controlled pinning/depinning, and these regimes should not be inferred from an exponent alone [60–63, 65]. Mixed or scale-dependent event populations can further produce similar statistical signatures from physically different deformation pathways [23, 56]. Consequently, meaningful comparison of reported exponents requires consideration of the active mechanism, event definition and experimental scale rather than the numerical exponent alone [3, 56].

## Experimental requirements for the next stage

The most useful next step is therefore not only to collect more stress–strain curves but to combine complementary observables in the same experiment. Simultaneous AE, DIC or local extensometry, diffraction and post-mortem microscopy would allow individual regimes to be connected to active carriers, localization length scales and evolving microstructure. Particularly valuable would be studies that compare compositionally related alloys, single-phase and multiphase states, different degrees of SRO, and room temperature versus elevated-temperature deformation under otherwise identical testing conditions. Future experiments should therefore use deliberately varied but well-documented thermo-mechanical histories within otherwise fixed compositions, so that the roles of processing, microstructure and chemical complexity can be separated rather than inferred from unrelated datasets.

If serration statistics are to be used prognostically, future experiments should explicitly test whether growth of the event cutoff or occurrence of an exceptionally large serration correlates reproducibly with damage initiation. Synchronizing stress drop statistics with AE source characterization, DIC localization, in situ imaging and post-mortem fracture analysis would allow a serration-based precursor to be distinguished from a large but non-damaging plastic avalanche [9, 59].

A more consistent reporting standard would greatly improve comparability. For AE, authors should specify threshold, hit-definition time, hit-lockout time, sensor band, sampling rate, coupling geometry and whether streamed or hit-based data were analyzed. For stress serrations, authors should state detrending procedures, machine stiffness if relevant, sampling frequency and whether the analyzed drops are type A, B, C or mixed. For DIC, spatial and temporal resolution, subset size and strain rate calculation method should be reported. These details are not secondary: in intermittent plasticity they define which part of the event hierarchy is visible.

From an engineering perspective, avalanche-like plasticity is neither intrinsically beneficial nor detrimental. Large synchronized events may reduce ductility, degrade surface quality or trigger strain localization. Distributed small avalanches may instead contribute to hardening and damage tolerance by spreading plasticity. The design challenge is therefore not to eliminate intermittency, which is intrinsic to defect-mediated deformation, but to control its scale and synchronization. CCAs are promising in this respect because chemistry, SRO, phase stability and microstructure can be tuned in high-dimensional composition space.

## Conclusions

Plastic deformation in CCAs cannot be understood only from the average shape of the stress–strain curve. The evidence reviewed here shows that apparently smooth flow, microplastic activity, AE bursts, localized strain bands and macroscopic serrations are different manifestations of intermittent defect dynamics observed at different scales. A central conclusion is therefore that CCA plasticity should be described not only in terms of strength, ductility and work hardening, but also in terms of how plastic strain is partitioned among discrete events and how these events become arrested, redistributed or synchronized.

The comparison with simple crystals and conventional alloys provides the necessary reference frame. Dislocation avalanches, microplastic bursts and PLC instabilities are not unique to CCAs. What makes CCAs especially interesting is that several sources of disorder and metastability coexist in the same material: chemical fluctuations, local lattice resistance, SRO, solute atmospheres, stacking fault energy variations, phase boundaries, precipitates and transformation or twinning pathways. These features can either suppress large avalanches by fragmenting slip or promote collective instabilities by storing elastic energy and releasing it abruptly. CCAs therefore offer a tunable platform for studying how microscopic defect motion becomes coupled to mesoscopic localization and macroscopic flow instability.

The CCA designation itself is therefore not a deformation mechanism or a sufficient predictor of intermittency. The active deformation carrier and its degree of synchronization are controlled more directly by crystal structure, stacking fault energy, phase stability, microstructure and testing conditions, while chemical complexity modifies the corresponding local energetic and kinetic landscapes.

The reviewed literature also shows that DSA is an important, but not universal, explanation for CCA intermittency. Elevated-temperature serrated flow in several systems is naturally interpreted within the DSA/PLC framework, particularly when negative strain-rate sensitivity, critical strain behavior or band propagation are observed. However, intermittent responses at room temperature, cryogenic temperature or during microplastic deformation may arise from other mechanisms, including source-limited plasticity, dislocation avalanches, twinning, transformation-assisted deformation, phase boundary interactions or adiabatic effects. The same mechanical signature can therefore have different microscopic origins, and the same microscopic process can appear differently depending on the measurement scale.

A broader implication is that CCA intermittency should be treated as a scale-bridging problem. Stress serrations are macroscopic, coarse-grained signals; AE detects faster and smaller cooperative events; local strain measurements reveal band dynamics; diffraction and microscopy identify the active structural carriers. No single observable is sufficient to determine the mechanism or the statistical character of deformation. Progress will require experiments that combine these methods, together with transparent statistical protocols and carefully controlled comparisons between alloy chemistry, processing state, phase constitution and temperature regime.

In this sense, avalanche-like plasticity is not merely a complication of mechanical testing. It provides a sensitive probe of how chemical complexity and microstructure control defect dynamics in CCAs. Understanding this intermittency may help connect alloy design to deformation stability, damage tolerance and reliable mechanical performance across temperatures and length scales. The challenge for future work is to move from documenting serrations or fitting power laws toward identifying the microstructural conditions under which intermittent plasticity is suppressed, amplified or organized into useful forms of strain accommodation. Controlled processing series studies will be essential for converting the present mechanism-oriented map into quantitative links between alloy design, microstructural state and avalanche statistics.

## Acknowledgements

This work received funding from the Czech Science Foundation, project No. 25-16368S and from the Operational Programme Johannes Amos Comenius of the MEYS of the Czech Republic, project Ferroic Multifunctionalities (FerrMion), No. CZ.02.01.01/00/22_008/0004591, co-funded by the European Union.

## Funding

Grantová Agentura České Republiky,25-16368S,Michal Knapek ,Ministerstvo Školství,Mládeže a Tělovýchovy ,CZ.02.01.01/00/22_008/0004591,Miloš Janeček

## Data availability

No new data were generated or analyzed in this review article. All data discussed are available in the cited literature.

## Declarations

**Conflict of interest** The authors declare no conflicts of interest.